\documentclass[12pt,preprint]{aastex}

\begin{document}

\title{The soft X-ray emission in a large
sample of galaxy clusters with ROSAT PSPC}

\author{Massimiliano~Bonamente$\,^{1,2}$, Richard~Lieu$\,^{1}$, 
Marshall K. Joy$\,^{2}$ and Jukka H. Nevalainen$\,^{1}$
}

\affil{\(^{\scriptstyle 1} \){Department of Physics, University of Alabama,
Huntsville, AL}\\
\(^{\scriptstyle 2} \){NASA Marshall Space Flight Center, Huntsville, AL}\\
}

\begin{abstract}
The study of soft X-ray emission of  38 X-ray selected galaxy clusters 
observed by {\it ROSAT} PSPC indicates that the {\it soft excess} phenomenon
may be a common occurrence in galaxy clusters. Excess soft X-ray radiation, above the contribution from
the hot intra-cluster medium, is evident in a large fraction of sources, and is clearly detected 
with large statistical significance in the deepest observations. 
The investigation relies on
new, high resolution 21 cm HI observations. 
The sample selection  also features analysis of infra-red images, to further ensure reliability
of results with respect to the characteristics of Galactic absorption.
The possibility of background or calibration effects as cause of the excess emission is likewise
investigated;  
a detailed analysis of the distribution 
of the excess emission with respect to detector position and
Galactic HI column density shows that the excess emission is a genuine
celestial phenomenon.
We find evidence for a preferential distribution of the soft excess emission at distances
larger than $\sim 150-200$ kpc from the centers of clusters; this behavior may be naturally explained
in the context of a non-thermal Inverse-Compton scenario. Alternatively, we propose that the phenomenon may
be caused by thermal emission of very large-scale `warm' filaments seen in
recent hydrodynamic simulations.  This new interpretation relieves the very demanding requirements
of either the traditional intra-cluster `warm' gas and the non-thermal scenarios.
We also investigate the possibility of the soft excess originating from unresolved, X-ray faint 
cluster galaxies.
\end{abstract}

\section{Introduction}
Clusters of galaxies are strong emitters of X-rays, which
originate from a very diffuse and hot phase of the intra-cluster medium (ICM).
At the typical temperatures of T $\sim$ 1-10 $\times 10^7$ K, the bulk
of the hot ICM radiation is detected near energies of  1  keV and above; intervening Galactic absorption
is responsible for a substantial reduction of flux
below $\sim$ 1 keV.

The soft X-ray band around 0.25 keV (hereafter C-band) offers a unique window
to investigate the presence of other emitting/absorbing phases
of the ICM. The initial discovery of soft X-ray and extreme ultra-violet (EUV) excess
emission in a few nearby galaxy clusters with the
{\it Extreme Ultra-Violet Explorer} and {\it ROSAT} missions
 (Coma, Virgo, A2199, A1795;  Lieu et al. 1996a,b;
Mittaz, Lieu and Lockman 1998; Lieu, Bonamente and Mittaz 1999)
was soon followed by many re-observations and re-analyses of the same objects, plus a
few more clusters, with the aid of complementary data from {\it BeppoSAX}, {\it Far Ultraviolet
Spectroscopic Explorer}, {\it Rossi X-Ray Timing Explorer} and {\it Hopkins Ultraviolet Telescope}
(Fabian 1996; Bonamente, Lieu and Mittaz 2001b; Bonamente et al. 2001c; Lieu et al. 1999b, 1999c;
Berghoefer, Bowyer and Korpela 2000a, 2000b; Bowyer and Berghoefer 1998; 
Bowyer and Berghoefer 2001; Bowyer, Berghoefer and Korpela 1999; 
Arabadjis and Bregman 1999; Reynolds et al. 1999; Kaastra
et al. 1999; Dixon et al. 2001; Dixon, Hurwitz and Ferguson 1996; 
Valinia et al. 2000; Fusco-Femiano et al. 2000; Buote 2000a,2000b,20001; Arnaud et al. 2001). 
The present analysis of a very large 
sample of X-ray selected clusters has the primary aim of assessing the
cosmological importance of this {\it soft excess} emission.
It is also known that some clusters may harbor, at their centers,
colder gas which absorbs the soft X-ray radiation (Fabian 1988). Recent
XMM observations  however did not detect large amounts of cold gas
in the center of galaxy clusters (e.g., Boehringer et al. 2002),
and our analysis of PSPC data shows
soft X-ray absorption only in the minority of the sources.

The PSPC detector on board {\it ROSAT} is perfectly suited for
studies of this kind. Along with a sizable effective area of $\sim$ 150 cm$^2$ at 0.25 keV,
pulse invariant (PI) channels 20-41 ($\sim$ 0.2-0.4 by photon energy)
covering the C-band are stable and well calibrated (e.g., Snowden et al. 1994).
The instrument, moreover, has a very low detector background, and
conducted a large number of deep cluster observations throughout its
lifetime. The large field of view of $\sim $ 1 degree radius permits also the
use of {\it in situ} background measurements, necessary for the
correct estimation and subtraction of the C-band background emission.

\section{Sample selection}
The sample of galaxy clusters we analyzed  
consists of the 38 clusters in Table 1, selected from
all the clusters observed by the PSPC during its pointed phase, 
according to the following criteria:
\begin{itemize}
\item they are located at high Galactic latitude ($\beta \geq$ 30 degrees) and with a count
rate $\geq$ 0.2 counts s$^{-1}$ in 0.5-2.0 keV band
(e.g., Schwope et al. 2000). The high Galactic latitude
portion of the sky
avoids regions of strong $N_H$ gradients along the Galactic plane;
\item Galactic HI column density $N_H \leq 5 \times 10^{20}$ cm$^{-2}$, {\it and} availability
of narrow beam HI measurement with resolution of $\sim$ 20 arcmin (Murphy et al., in prep.).
The measurements of Murphy et al. (in prep.) with the 140-ft NRAO 
telescope at Green Bank constitute the best Galactic $N_H$ data
toward galaxy clusters to date (the Dickey and
Lockman survey [Dickey and Lockman, 1990] has a resolution of $\sim$1
degree), with a beam-size  comparable to the X-ray extent of the
clusters;
\item analysis of IRAS 100 $\mu m$ images (Wheelock et al. 1994) confirms smoothness of the
Galactic absorbing gas distribution. It is in fact known
that the 100 $\mu m$ emission correlates well with $N_H$,
and IRAS maps can further detect asymmetries and gradients in the
distribution of the Galactic absorbing material with
a better resolution of approximately 3 arcmin.
\end{itemize}
With these conditions in place, the $N_H$ toward the clusters, a
primary parameter that ensures  the accurate reconstruction of C-band
signals, can be known with an uncertainty of less than $1 \times 10^{19}$
cm$^{-2}$ (Murphy et al., in prep.). Moreover, given that through a column density of
 4 $\times 10^{20}$ cm$^{-2}$ a 0.25 keV photon has an optical depth of $\sim$ 4, we
restricted our analysis to sources with $N_H \leq 5 \times 10^{20}$
cm$^{-2}$.
Many rich and
nearby clusters (e.g., A496, A4038, A2052 etc.) were not
considered in this sample, because either a narrow beam $N_H$
measurement was unavailable, or on account of
their positional coincidence with a Galactic IR cirrus gradient. In addition,
all clusters~\footnote{With the only possible exceptions of Coma and Virgo,
where marginal cluster emission may be present even near the
detector edges. In these two clusters, analyzed only to a radius of
18 arcmin to minimize background effects,
the background is estimated from the outermost detector regions,
and amounts to $\leq$ 20\%
of the cluster signals at most. Even a $\sim$ 50\% uncertainty on the background level results
in a $\leq$ 10\% effect on the soft excess fluxes, although the
background is likely an overestimate which, in turn,
underestimates the soft excess fluxes.} have an X-ray angular extent of
less than 1 degree radius, and local background can be obtained
from the outer regions of the PSPC detector.

\begin{deluxetable}{lccclc}
\tablecaption{Characteristics of galaxy clusters analyzed in this paper}
\tablehead{\colhead{Cluster} & \colhead{$z$} & \colhead{$N_H$  ($10^{20}$ cm$^{-2}$)} & 
\colhead{CF} & \colhead{$r_c$ (kpc)} & \colhead{Abell  richness}}
\startdata
A21 & 0.095 & 3.7 & no [WJF97] & 600 [W97] & 1  \\
A85 & 0.055 & 2.7 & yes [WJF97] & 317 [M99] & 1   \\
A133 &  0.057 & 1.5 & yes [WJF97] & 300 [W97] & 0  \\
A665 & 0.1816 & 4.3 &  no [WJF97] & 1000 [W97] & 5  \\
A1045 &0.138 & 1.4 & \nodata & 240 [c] &0  \\
A1068 & 0.139 & 1.57 & yes [A00] & 390 [E96] &0  \\
A1302 & 0.116 & 1.0 &\nodata  & 330 [E96] & 0  \\
A1314 & 0.0341 & 1.6 & no [WJF97] & 500 [W97] &0  \\
A1361 & 0.1167 & 2.5 & yes [A95]& 270 [E96] &0  \\
A1367 & 0.0276 & 1.7 & unc. [WFJ97,P98] & 360 [M99] & 2 \\
A1413 & 0.143 & 1.71 & no [WJF97]& 500 [W97] & 3 \\
A1689 & 0.181 & 1.72 & no [WJF97,P98]& 131 [M99] & 4 \\
A1795 & 0.061 & 1.01 & yes [WJF97]& 344 [M99]& 2 \\
A1914 & 0.171 & 0.95 &\nodata  & 580 [E96] &2 \\
A1991 & 0.0586 & 2.25 & yes [WJF97] &  200 [W97] & 1 \\
A2029 & 0.0767 & 3.2 & yes [WJF97]& 334 [M99] &2  \\
A2142 & 0.09 & 4.15 & yes [WJF97] & 580 [HB96] &2  \\
A2199 & 0.0302 & 0.85 & yes [WJF97] & 162 [M99] & 2  \\
A2218 & 0.171 & 2.75 & unc. [WJF97,A00] & 230 [NB99] &4 \\
A2219 & 0.228 & 2.04 & no [A00]& 650 [E96] & 0 \\
A2241 & 0.0635 & 2.05 &\nodata  & 80 [c] &0 \\
A2244 & 0.097 & 1.9 & yes [WJF97]& 117 [M99] &2 \\
A2255 & 0.08 & 2.51 & no [WJF97]& 584 [M99] & 2 \\
A2256 & 0.06 & 4.55 & no [WJF97]& 486 [M99] & 2 \\
A2597 & 0.085 & 2.2 & yes [A01]& 44 [M99] &0  \\
A2670 & 0.076 & 2.73 & yes [WJF97]& 600 [W97]&3  \\
A2717 & 0.049 & 1.23 & \nodata    & 47 [L97] & 1 \\
A2744 & 0.308 & 1.41 & no [A00]& 1000 [E96] & 3 \\
A3301 & 0.054 & 2.49 &  &  140 [c] &3  \\
A3558 & 0.048 & 4.0 & yes [P98]& 194 [M99] &4 \\
A3560 & 0.04 &4.7  & \nodata &  170 [E96] &3  \\
A3562 & 0.04 &4.01  & no [WJF97]& 97 [M99] &2  \\
A3571 & 0.04 & 4.4 & yes [P98]& 173 [M99]& 2 \\
A4059 & 0.046 & 1.06 & yes [WJF97] & 78 [M99]& 1 \\
Coma (A1656) & 0.023 & 0.9  & no [WJF97]& 386 [M99] & 2 \\
Fornax (AS373) & 0.0046 & 1.5 & yes [WJF97] & 10 [W97] & 0 \\
Hercules (A2151) & 0.037 & 3.2 & yes [WJF97]& 35 [HB99] & 2 \\
Virgo (M87) & 0.0043 & 1.8 & yes & 300 [W97] & \nodata \\
\enddata
\tablecomments{Clusters are ordered alphabetically,
$z$ is redshift, $N_H$ is Galactic HI column density, CF indicates whether the source is believed to host cooler gas at its center,
$r_c$ is X-ray core radius.\\
References:
WJF97 and W97 - White, Jones and Forman 1997; A00 - Allen 2000;
A01 - Allen, Fabian, Johnstone, Arnaud and Nulsen 2001;
A95 - Allen, Fabian, Edge, Boheringer and White 1995;
P98 - Peres, Fabian, Edge, Allen, Johnston and White 1998;
E96 - Ebeling et al. 1996;
M99 - Mohr et al. 1999;
HB96 - Henry and Briel 1996 (A2142) ;
L97  - Liang et al. 1997 (2717); 
HS96 - Huang and Sarazin 1996 (Hercules) ;
NB99 - Neumann and Boheringer 1999 (A2218) ;
c - this work}
\end{deluxetable}

\begin{deluxetable}{lccc}
\tablecaption{\footnotesize Log of PSPC observations. Nearly 570 ks of PSPC data was reduced and analyzed.}
\tablehead{\colhead{Cluster} &\colhead{Obs. ID}&\colhead{Obs. date}& \colhead{Exp. time (ks)}}
\startdata
\hline
A21 & 800107 & 25/12/91 & 8.5 \\
A85 & 800250 & 1/7/92 & 8.4 \\
   & 800174a00 & 20/12/91 & 2.1 \\
   & 800174a01 & 11/6/92 & 3.3 \\
A133 & 800319 & 5/7/92 & 15.5 \\
A665 & 800022 & 10/4/91 & 35.0 \\
A1045 & 800350 & 28/5/93 & 9.1 \\
A1068 & 800410 & 4/12/92 & 9.8 \\
A1302 & 800260 & 29/9/92 & 2.3 \\
A1314 & 800392 & 9/6/93 & 2.0 \\
A1361 & 800412 & 7/6/93 & 3.9 \\
A1367 & 800153 & 2/12/91 & 17.8 \\
A1413 & 800183 & 27/11/91 & 14.3 \\
A1689 & 800248 & 18/7/92 & 13.5 \\
A1795 & 800055 & 1/7/91 & 25.0 \\
      & 800105 & 4/1/92 & 35.0 \\
A1914 & 800345 & 20/7/92 & 6.4 \\
A1991 & 800518a01 & 4/8/93 & 20.9 \\
      & 800294 & 8/8/92 & 2.8 \\
A2029 & 800249 & 10/8/92 & 9.9 \\
A2142 & 150084 & 20/7/90 & 7.5 \\
      & 800096 & 25/8/92 & 5.9 \\
      & 800233 & 26/8/92 & 4.8 \\
A2199 & 150083 & 18/7/90 & 10.0 \\
A2218 & 800097 & 25/5/91 & 36.4 \\
A2219 & 800571 & 2/8/93 & 8.5 \\
A2241 & 800530 & 3/8/93 & 9.9 \\
A2244 & 800265 & 21/9/92 & 2.9 \\
A2255 & 800512 & 5/9/93 & 14.5 \\
A2256 & 100110 & 17/6/90 & 16.6 \\
A2597 & 800112 & 27/11/91 & 6.7 \\
A2670 & 800420 & 29/6/92 & 14.7 \\
A2717 & 800347 & 23/5/93 & 8.8 \\
A2744 & 800343 & 16/6/92 & 13.6 \\
A3301 & 800391 & 22/9/92 & 8.3 \\
A3558 & 800076 & 17/7/91 & 28.0 \\
A3560 & 800381 & 9/8/92 & 5.3 \\
A3562 & 800237 & 19/1/93 & 17.6 \\
A3571 & 800287 & 12/8/92 & 5.1 \\
A4059 & 800175 & 21/1/91 & 5.3 \\
Coma (A1656)& 800005 & 17/6/91 & 20.8 \\
Fornax (AS373) & 600043 & 15/8/91 & 52.8 \\
Hercules (A2151) & 800517 & 17/8/93 & 11.0 \\
Virgo & 800187 & 10/6/92 & 9.4 \\
\enddata
\end{deluxetable}

\section{Data and data analysis}
The PSPC data were reduced according to the prescriptions of Snowden
et al. (1994). The datasets, available through public archives
(NASA's W3Browse and the Leicester LEDAS databases) were corrected
for detector gain fluctuations, and only events with Average Master Veto
rate $\leq$ 170 c s$^{-1}$ were considered, in order to discard periods of
high particle background. 
The PSPC rejection efficiency for particle background is 99.9\% in the energy range 0.2-2 keV (Plucinsky et al. 1993),
and the background is therefore solely represented by the photonic component.
After a `cleaned' dataset was obtained,
each observation was divided into a number of concentric annular regions between 2 and 8,
according to the size and the brightness of each source; the linear size of each region
was never less than 1 arcmin.
Response files associated with the spectra account for the instrument's point-spread
function (PSF) and
vignetting. Moreover, all obvious point sources were accurately
removed.

The C-band background is dominated by the Galactic soft X-ray emission
 (e.g., Snowden et al. 1998). Owing to the large field of
view of the PSPC, it is possible to measure  the background from a
peripheral detector region (typically an annular region near 40 arcmin from
boresight) where cluster emission is negligible. A background
spectrum is extracted for each dataset, and
rescaled to each cluster region according to
area, vignetting  and the energy-dependent detector response at
different detector positions (see also Bonamente et al. 2001c).
This {\it in situ} background is the most accurate for the purpose
of background subtraction, as it is taken at the same time as the
cluster spectra, and at a minimal angular  distance from the cluster's
position.
Peripheral cluster regions are most affected by the background;
regions where the cluster emission drops below 25\% of the
background can be
considered as limiting radii of the C-band detection for each cluster.
The integrity of the employed background subtraction technique is implied
by the absence of excess emission/absorption in those regions
where negligible cluster signal is detected (see next section).

Dedicated narrow-beam 21-cm observations are available from the
NRAO 140-ft Green Bank telescope, with a resolution of $\sim$ 20
arcmin.  They provide $N_H$ measurements
with an accuracy of less than $1 \times 10^{19}$ cm$^{-2}$, (Murphy et al., in prep.; see
Table 2) for a large number of galaxy clusters.
On occasions for which a comparison with stellar Ly$\alpha$ and quasar
X-ray spectra could be made, the agreement among the different methods of determining
$N_H$ would again suggest  an error of $\leq 10^{19}$ cm$^{-2}$ (see 
Laor et al. 1994; Elvis, Lockman  and Wilkes 1989; Dickey and Lockman 1990), 
irrelevant to the results of this study. 
In addition, IRAS 100 $\mu$m maps (with a resolution of $\sim$ 3 arcmin)
can give further
constraints on the state and distribution of the Galactic
absorbing material.
In accordance with the $N_H$-IR correlation (with scaling factor of
$\sim 1.2 \times 10^{20}$ cm$^{-2}$ (MJy sr$^{-1}$)$^{-1}$,
Boulanger and Perault 1988), we checked that no strong IR
gradients were present in a 1 $\times$ 1 degree IRAS field around
the cluster, and that quantitatively the $N_H$ toward the
background region was consistent (to within a few $10^{19}$
cm$^{-2}$) with that measured at the cluster position.
When this check failed, the candidate was discarded from the
sample. On the basis of this criterion, A496 and A500 are not
considered here, as they both lay in a region of strong Galactic
IR cirrus gradient.

The limited band and energy resolution of the PSPC instrument (0.2-2 keV) is not
ideally suited for the determination of the hot ICM parameters (temperature T and
metal abundance A) of
very hot clusters (T $\geq$ 3-4 keV), although it is proven that deep PSPC observations can
yield accurate temperature maps even for hot clusters (e.g., Henry and Briel 1996).
To avoid the controversy, we employed the spatially resolved best-fit hot 
ICM parameters available elsewhere in the
literature where kT $\geq 3-4$ keV (see Table 3 for references);
with kT and A fixed at their best fit, and $N_H$ at the
measured values, we fitted the PSPC 0.5-2 keV spectra to
a photoabsorbed thin plasma emission model. The goodness of fit
 implies that 0.5-2 keV PSPC spectra are well described
by the employed models.
For those cluster regions where kT $\leq$ 3-4 keV, or those cases where literature values were
unavailable, we fitted
the hot ICM parameters to the 0.5-2 keV PSPC spectra.
The reason for avoiding the 0.2-0.5 keV portion of the PSPC
spectra to model the hot ICM emission  is that such an undertaking
enables us to study the C-band emission in an unbiased fashion. It is clear that, e.g.,
any excess emission in 0.2-0.4 keV band will result
in a tendency of the hot ICM parameters toward lower kT,
if the entire
0.2-2 keV spectrum is used for the fit of the hot ICM 
model~\footnote{This
is in fact evident in the case of S\'{e}rsic 159-03, see Bonamente, Lieu and Mittaz
(2001d), where the strong soft excess prevents
a good fit to be found over the whole 0.2-2 keV range, and the best-fit parameters to be discordant
 with those measured by EPIC/XMM in a slightly higher energy band.}.

The optically thin emission code is MEKAL (in XSPEC language;
Mewe, Gronenshild and van den Oord 1985; Mewe, Lemen and van den Oord 1986; Kaastra 1992), 
and Galactic absorption was modeled with the codes of
Morrison and McCammon 1983 (WABS). The cross sections of WABS are in
good agreement with the recent compilation of Yan, Saghdepour and
Dalgarno (1998), as also shown in Bonamente, Lieu and Mittaz (2001b). The
most up-to-date code by Wilms, Allen and McCray 2000 (TBABS) is
virtually indistinguishable from WABS at the resolution of PSPC,
and yield the same results (see Bonamente, Lieu and Mittaz 2001d).

Once a model of the hot ICM is obtained, detected C-band fluxes are
compared with model predictions, see Table 3.
The fractional excess/absorption function $\eta$ is defined as
\begin{equation}
 \eta = \frac{p-q}{q} 
\end{equation}
where $p$ is detected C-band flux, and $q$ is the C-band model prediction
according to the thermal models of Table 3.
A value of, e.g., $\eta$=0.2 in the fractional excess plots indicates that C-band
emission is detected at the level of
20\% excess with respect to the hot ICM emission at that
cluster position.
The effects of any gradients in the temperature (e.g., De Grandi and Molendi 2002)
and abundance (e.g., De Grandi and Molendi 2001) distribution with
cluster radius on the soft excess fluxes is accounted for by the spatially
resolved modeling of the hot ICM. 
%Uncertainties in the hot ICM parameters are also considered:
%the errors in $\eta$ are in fact obtained by proper addition of the
%statistical uncertainties in C-band fluxes and hot ICM model
%uncertainties. The latter is estimated from the statistical error
%in the 0.5-2 keV PSPC data (the portion of the spectra used for
%model determination), and normally contributes to only a small
%fraction of the total $\Delta \eta$.
Uncertainties in the hot ICM parameters are also considered. The total errors
$\Delta \eta$ are in fact obtained by proper addition of
statistical uncertainties in C-band fluxes ($\Delta$p) and statistical
uncertainties in the hot ICM models ($\Delta$q). The latter is
calculated as the sum of model variations when the best-fit T is varied by 
$\Delta T = \pm 10$\% T ($\Delta$ q$_T$) and the best-fit
A by $\Delta A = \pm 30$\% A ($\Delta$ q$_A$); this choice 
is consistent with the published hot ICM model uncertainties
for the clusters in this sample (see Table 3 for references), and it is
oftentimes a conservative estimate (e.g., as in the cases of Coma or Virgo, where temperatures
are known with accuracy of $\sim 1-2$\%).
 This method of estimation
of hot ICM model uncertainties is therefore uniform throughout the
sample, and does not depend on the quality of the PSPC data or of the
other data used to determine the best-fit ICM models. 

\begin{deluxetable}{lcccccc}
\tabletypesize{\small}
\tablecaption{Spatially resolved best-fit models of the hot 
ICM and C-band fluxes for all clusters. }
\tablehead{
Region & kT   & A & $\chi^2_{r}$ & C-band flux & $L_{NT}$ & $L_{therm}$ \\
(arcmin)  &  (keV)      &   & &  $10^{-2}$c s$^{-1}$ & $10^{41}$ erg s$^{-1}$ &$10^{41}$ erg s$^{-1}$}
%\colhead{Region} & \colhead{kT  keV} & \colhead{red. $\chi^2$ (dof)} &
%\colhead{C-band flux (model) $10^{-2}$c s$^{-1}$} & \colhead{$L_{NT}$ $10^{41}$ erg s$^{-1}$} &
%\colhead{$L_{therm}$  $10^{41}$ erg s$^{-1}$}}
\startdata
\multicolumn{7}{c}{{\bf A21} (z=0.055, $N_H=3.7$, Ref: [E96])} \\
0-1 & 4 & 0.4  & 0.71(22) & 0.57$\pm$0.1 (0.58) &\nodata & \nodata  \\
1-3 & 4$\pm^{5.5}_{1.6}$ & 0.3  & 0.88(47) & 1.25$\pm$0.16 (1.33)& \nodata &\nodata \\
3-6 & 6.5$\pm^{}_{3.7}$ & 0.3  & 1.17(51) & 1.2$\pm$0.26 (1.1)&\nodata &\nodata \\
6-9 & 5 & 0.3 &  0.71(39) & 1$\pm$0.33 (0.51) (16\%) &\nodata & \nodata  \\
9-12 & 5 & 0.3 &  0.95(38) & 0.16$\pm$0.2 (0.19) (2\%) &\nodata &\nodata \\
\multicolumn{7}{c}{{\bf A85} (z=0.055, $N_H$=2.7, Ref:[W00,L01,MFSV98])} \\
 0-1 & 3.4$\pm^{1}_{0.7}$ & 0.5 & 1.12(130) & 9.8$\pm$0.35 (10.68) &\nodata &\nodata \\
 1-3 & 5.1$\pm^{1.9}_{1.2}$& 0.5& 1.09(139)& 14.6$\pm$0.43 (13.47)& 2.3$\pm^{85}_{2.3}$& 6.5$\pm^{5}_{6.5}$ \\
 3-6 & 7 & 0.3  & 1.13(138)& 12.0$\pm$0.35 (11.7) & \nodata & \nodata\\
 6-9 & 7 & 0.3  & 0.89(123) & 7.6$\pm$0.28 (6.5) & 9.1$\pm^{15}_{9.1}$ & 3.8$\pm^{5}_{3.8}$ \\
 9-12& 7 & 0.3 & 0.91(101) & 4.7$\pm$0.26 (4.1) & 1.3$\pm^{15}_{1.3}$ & 1.4$\pm^{6}_{1.4}$ \\
12-15& 7 & 0.3 & 0.99(73) & 2.1$\pm$0.3 (2.2) (15\%)&\nodata &\nodata  \\
\multicolumn{7}{c}{{\bf A133} (z=0.057, $N_H$=1.5, Ref:[R81,I02])} \\
0-1 & 2.6$\pm^{0.4}_{0.4}$ & 1.6$\pm^{1.5}_{0.5}$ & 1.24(119) & 8.2$\pm$ 0.23 (6.5) &14$\pm{0.4}$ & 7.5$\pm^{2.4}_{2.2}$ \\
1-3 & 4.1$\pm^{1.6}_{1}$ & 0.44$\pm^{0.72}_{0.37}$  & 0.88(123) & 8.8$\pm$ 0.25 (7.9) &8.2$\pm^{8}_{6.5}$ & 3.2 $\pm{2.1}$  \\
3-6 & 5.7$\pm^{5.5}_{2}$ & 0.3 & 1.11(111) & 6.6$\pm$ 0.26 (5.3) & 4.6$\pm^{8.5}_{4.6}$ & 6.5$\pm^{2.3}_{2.1}$  \\
6-9 & 5 & 0.3 & 0.91(77) & 2.6$\pm$0.24 (2.4) &\nodata &\nodata \\
9-12 & 5& 0.3  & 1.09 (56) & 1$\pm$0.25 (1.17) (11\%) &\nodata &\nodata  \\
\multicolumn{7}{c}{{\bf A665} (z=0.186, $N_H$=4.3, Ref:[W00])} \\
0-1 & 5$\pm^{1.8}_{1.1}$ & 0.3  & 1.14(122) & 9.6$\pm$0.58 (11.6) &\nodata &\nodata \\
1-3 & 7 & 0.3 & 0.86(131) & 1.43$\pm$0.09 (1.41) &\nodata &\nodata \\
3-6 & 7 & 0.3 & 1.02 (126) & 0.8$\pm$0.13 (0.91) (16\%) &\nodata &\nodata \\
\multicolumn{7}{c}{{\bf A1045} (z=0.138, $N_H$=1.4)}  \\
0-1 & 3.8$\pm^{3.7}_{1.5}$ & 0.3  & 0.86(43) & 2.7$\pm$ 0.18 (3.3) &\nodata & \nodata \\
1-3 & 4.5$\pm^{}_{2.1}$ & 0.3  & 0.7(28) & 1.86$\pm$0.21 (1.89) & \nodata&\nodata \\
3-6 & 8.8$\pm^{}_{6.3} $ & 0.3 & 0.36(20) & 0.43$\pm$0.32 (0.78) (6\%) &\nodata &\nodata  \\
\multicolumn{7}{c}{{\bf A1068} (z=0.139, $N_H$=1.57, Ref:[W00])} \\
0-1 & 3.3$\pm^{1.6}_{0.9}$ & 0.3  & 1.28(90) & 8.4$\pm$0.3 (8.6) &\nodata &\nodata \\
1-3 & 5 & 0.3  & 0.71(45) & 34$\pm$0.23 (27.5) & 55$\pm{34}$ & 24$\pm{13}$\\
3-6 & 5 & 0.3  & 1.02(24) & 0.6$\pm$0.28 (1) (10\%) &\nodata &\nodata \\
\multicolumn{7}{c}{{\bf A1302} (z=0.116, $N_H$=1.0, Ref:[E96])} \\
0-1 & 4.8 & 0.3  & 1.33(8) & 2.63$\pm$0.35 (2.83) &\nodata &\nodata \\
1-3 & 4.8  & 0.3   & 0.99(10) & 4$\pm$0.5 (3.4) & 9.3$\pm^{38.5}_{9.3}$& 3.1$\pm^{12}_{2.8}$\\
3-6 & 4.8  & 0.3  & 0.30(7) & 2.4$\pm$0.62 (1.8) &\nodata &\nodata \\
6-9 & 2.4$\pm^{}_{1.4}$ & 0.3 & 0.31(3) & 0.8$\pm$0.7 (0.9) (10\%) &\nodata &\nodata \\
\multicolumn{7}{c}{{\bf A1314} (z=0.034, $N_H$=1.6, Ref:[E96])} \\
0-6 & 5$\pm^{}_{3.2}$ & 0.3  & 0.71(15) & 5.33$\pm$0.8 (3.7) & 5.6$\pm^{6.9}_{5.6}$& 2.5$\pm^{2.3}_{2.5}$ \\
6-9 & 5   & 0.3 & 0.72(13) & 3.5$\pm$0.9 (1.77)  & 6.2$\pm^{7.8}_{6.2}$ & 2.8$\pm^{4.2}_{2.8}$\\
9-12 & 5 & 0.3  & 0.57(10) & 0.3$\pm$0.9 (1.4) (3\%) &\nodata &\nodata  \\
\multicolumn{7}{c}{{\bf A1361} (z=0.116, $N_H$=2.5, Ref:[E96])} \\
0-1 & 2.2$\pm^{1.4}_{0.5}$ & 0.54$\pm^{2}_{0.4}$ & 0.86(27) & 2.44$\pm$0.26 (2.82) &\nodata  &\nodata\\
1-3 & 2.9$\pm^{7.1}_{1.2}$ & 0.3 &  0.67(14) & 1.77$\pm$0.28 (1.5) & \nodata&\nodata \\
3-6 & 5 & 0.3 &  1.4(8) & 0.64$\pm$0.4 (0.5) (14\%) &\nodata &\nodata \\
\multicolumn{7}{c}{{\bf A1367} (z=0.0277, $N_H$=1.7, Ref:[I02])} \\
0-3 & 3.4$\pm^{2.2}_{1.2}$ & 0.3 & 1.39(113) & 4.8$\pm$0.19 (4.83) &\nodata &\nodata \\
3-6 & 2.5$\pm^{0.5}_{0.35}$ & 0.3 & 0.99(134) & 10.3$\pm$0.3 (10.17) & \nodata&\nodata\\
6-9 & 3.6 $\pm^{1.1}_{0.7}$ & 0.3 & 0.94(136) & 10.26$\pm$ 0.33 (10.07) & \nodata& \nodata\\
9-12 & 3.2 & 0.3 &  0.98(137)& 9.8$\pm$0.36 (9.4) &\nodata &\nodata \\
12-15 & 3.2 & 0.3 &  1.06(134) & 7.85$\pm$0.37 (7.96) (37\%) &\nodata &\nodata \\
\multicolumn{7}{c}{{\bf A1413} (z=0.143, $N_H$=1.71, Ref:[W00])} \\
0-1 & 8 & 0.4 &  0.79(69) & 5.4$\pm$0.28 (5.63) & &\\
1-3 & 6 & 0.3 &  1.07(72) & 6.5$\pm$0.3 (5.85) & 5$\pm^{40}_{5}$& 16$\pm^{21}_{16}$ \\
3-6 & 6 & 0.3 &  1.2(36) & 2.39$\pm$0.31 (2.1) &\nodata &\nodata\\
6-9 & 6 & 0.3 &  0.84 (22) & 0.26$\pm$0.33 (0.69)(3.5\%) & \nodata&\nodata\\
\multicolumn{7}{c}{{\bf A1689} (z=0.181, $N_H$=1.72, Ref:[W00])} \\
0-1 & 5.1$\pm^{1.31}_{1.1}$ & 0.5  & 1.25(124) & 8.46$\pm$0.25 (9) &\nodata &\nodata \\
1-3 & 8 & 0.3 &  1.04(107) & 5.27$\pm$0.226 (5) & 0.7$\pm^{16}_{0.7}$ & 8.6$\pm^{27}_{8.6}$ \\
3-6 & 7 & 0.3 &  0.85(52) & 1.57$\pm$0.23 (1.54) (23\%)&\nodata &\nodata\\
\multicolumn{7}{c}{{\bf A1795} (z=0.063, $N_H$=10.1, Ref:[BH96])} \\
0-1 &3.6$\pm^{0.3}_{0.2}$ & 0.73$\pm^{0.15}_{0.12}$ &1.04(301) & 35.0$\pm$0.28 (32.83)& 17$\pm{5.3}$ & 7$\pm{1.6}$ \\
1-2 &6.4$\pm^{1}_{0.8}$ & 0.58$\pm^{0.42}_{0.37}$ & 1.14(300) & 23.9$\pm$0.24 (20.22) & 34$\pm^{4.5}_{2.8}$ & 15$\pm{1.4}$ \\
2-3 &6.2$\pm^{1.3}_{1}$ & 0.56$\pm^{0.66}_{0.42}$ & 1.27(287) & 14.1$\pm$0.19 (12.2) & 17$\pm{2.9}$ & 7.3$\pm{1.1}$\\
3-6 &7 & 0.3 & 1.13(290)& 21.7$\pm$0.22 (19.53) &  19$\pm{4.6}$& 8.3$\pm^{1.6}_{1.4}$\\
6-9 &7 & 0.3 & 0.97(278)& 8.05$\pm$0.15 (7.5) & 3.2$\pm^{3.4}_{3.2}$ & 1.7$\pm^{1.5}_{1.4}$\\
9-12&7 & 0.3& 0.85(244) & 3.44$\pm$0.15 (3.45) (19\%)&\nodata &\nodata\\
\multicolumn{7}{c}{{\bf A1914} (z=0.171, $N_H$=0.95, Ref:[W00])} \\
0-1 & 10 & 0.5 &  1.21(82) & 11.05$\pm$0.42 (11.1) & \nodata&\nodata\\
1-3 & 10 & 0.3 &  0.98(58) & 8.4$\pm$0.4 (7.7) & 4.7$\pm^{56}_{4.7}$ & 5.7$\pm^{22}_{5.7}$ \\
3-6 & 9 & 0.3 &   1(151) & 2.2$\pm$0.37 (2.33) (25\%) &\nodata &\nodata\\
6-9 & 8 & 0.3 &   0.7(22) & 0$\pm$0.71 (0.6) (0\%) &\nodata &\nodata\\
\multicolumn{7}{c}{{\bf A1991} (z=0.058, $N_H$=2.25, Ref:[E96])}\\
%0-1 & 1.7$\pm^{0.3}_{0.3}$ & 1.2$\pm^{1.7}_{0.65}$ & 0.98(32) & 13.25$\pm$0.69 (1.254) &\nodata &\nodata\\
%1-3 & 5 & 0.3 &  0.74(22) & 3$\pm$0.4 (3.3) &\nodata &\nodata\\
%3-6 & 5 & 0.3 &  0.51(15) & 1.65$\pm$0.49 (1.65) (25\%) &\nodata &\nodata\\
%6-9 & 4 & 0.3 &  1.3(11) & 1.6$\pm$0.6 (0.65) (17 \%) & \nodata&\nodata\\
%9-12 & 4 & 0.3 &  1.0(11) & 1.6$\pm$0.7 (2.5) (13 \%) &\nodata &\nodata\\
0-3 & 2.0$\pm^{0.2}_{0.18}$ & 0.78$\pm^{0.25}_{0.19}$ & 1.1(136) & 6.5$\pm$0.2 (7.16) &\nodata &\nodata \\
3-6 & 3.5$\pm^{2.7}_{1.15}$ & 0.65$\pm^{3.3}_{0.58}$ & 0.81(88) & 1.89$\pm$0.18 (1.59) & 0.53$\pm^{19}_{0.53}$ & 2$\pm^{6.3}_{2}$ \\
6-9 & 4  & 0.3 & 0.89(70) & 0.78$\pm$0.2 (0.51) (10\%) &\nodata &\nodata \\
\multicolumn{7}{c}{{\bf A2029} (z=0.077, $N_H$=3.2, Ref:[MFSV98])} \\
0-1 & 3.5$\pm^{0.8}_{0.4}$ & 0.5 &  1.04(144)& 13.23$\pm$ 0.37 (13.9) &\nodata &\nodata \\
1-3 & 9 & 0.3 &  1.07(145) & 13.36$\pm$0.387 (11.98) & 21$\pm^{41}_{21}$& 22$\pm^{15.5}_{13}$ \\
3-6 & 9 & 0.3 &  0.95(134) & 6.86$\pm$0.34 (6.24) & 1.6$\pm^{34}_{1.6}$ & 11$\pm^{13.5}_{11}$ \\
6-9 & 9 & 0.3 &  1.05(107) & 2.81$\pm$0.33 (2.72) &\nodata &\nodata\\
9-12& 9 & 0.3 &  0.85(82) & 0.09$\pm$0.35 (1.34) (8.2\%) &\nodata &\nodata\\
\multicolumn{7}{c}{{\bf A2142} (z=0.09, $N_H$=4.15, Ref:[HB96])}\\
0-1  & 5.8$\pm^{1.85}_{1.2}$ & 0.67$\pm^{1.1}_{0.6}$ & 0.89(149) & 5.69$\pm$0.2 (5.66)& \nodata& \nodata\\
1-3 & 9 & 0.3 &  0.75(151) & 7.93$\pm$0.2 (7.72) &\nodata &\nodata\\
3-6 & 9  &0.3 & 0.90(144) & 5.26$\pm$0.24 (4.87) &1.6$\pm^{37}_{1.6}$ & 12$\pm^{19.6}_{12}$\\
6-9 & 9 & 0.3& 0.69(136) & 2.16$\pm$0.24 (2.6) & \nodata & \nodata \\
9-12 &9  & 0.3  & 0.68(104) & 0.94$\pm$0.35 (1) &\nodata &\nodata\\
12-15 &9  & 0.3 & 0.60(92)  & 0.59$\pm$0.25 (0.65) &\nodata &\nodata\\
\multicolumn{7}{c}{{\bf A2199} (z=0.03, $N_H$=0.85, Ref:[MVFS99,SSJ98])} \\
0-1 & 3.2$\pm{0.7}$ & 0.4 &  1.19(138) & 28.2$\pm$0.54 (33.27) &\nodata &\nodata \\
1-2 & 4.3$\pm^{1.4}_{0.9}$ & 0.4 &  0.93(137) & 32.14$\pm$0.58 (30.53) & 0.05$\pm^{0.96}_{0.05}$ & 0.46$\pm^{0.7}_{0.46}$ \\
2-3 & 4.2$\pm^{1.7}_{1.1}$ & 0.3 &  1.09(129) & 22.61$\pm$ 0.5 (23.72) & \nodata& \nodata\\
3-6 & 4.5 & 0.3 &   1.05(146) & 46.94$\pm$0.76 (47.37) & \nodata & \nodata \\
6-9 & 4.5 & 0.3 & 0.92(130) & 24.53$\pm$0.63 (23.6) &0.5$\pm{0.5}$ &  0.22$\pm^{0.4}_{0.22}$\\
9-12& 4.5 &  0.3 & 1.03(114) & 12.55$\pm$0.57 (12.69) &\nodata &\nodata \\
12-15&4.5 &  0.3 & 0.95(90)  & 8.98$\pm$0.59 (7.18) & 4.8$\pm^{3.3}_{3.4}$&1.7$\pm^{1.1}_{1.15}$\\
15-18 & 4.5& 0.3 & 0.65(81) & 5.9$\pm$0.62 (5.2) (19\%) &\nodata &\nodata \\
\multicolumn{7}{c}{{\bf A2218} (z=0.17, $N_H$=2.75, Ref:[W00,NB99])} \\
0-1 & 5.4$\pm^{2.3}_{1.4}$ & 0.36$\pm^{1.95}_{0.36}$& 1.32(118) & 1.93$\pm$ 0.08 (1.77) &0.47$\pm^{61}_{0.47}$ & 12$\pm^{15.3}_{12}$ \\
1-3 & 7& 0.2 &  1.17(124) & 2.35$\pm$0.1 (1.93) & 59$\pm^{44}_{34.5}$& 29$\pm{19}$\\
3-6 & 6 & 0.2 & 0.88(103) & 1.1$\pm$0.14 (8.5) (16\%) &\nodata &\nodata\\
6-9 & 5 & 0.2 & 1.37(78) & 0.05$\pm$0.17 (0.28) (1\%) &\nodata &\nodata\\
\multicolumn{7}{c}{{\bf A2219} (z=0.23, $N_H$=2.04, Ref:[W00])} \\
0-1 & 6.6$\pm^{10}_{2.6}$ & 0.3 & 0.97(61) & 3.6$\pm$0.21 (3.7) &\nodata &\nodata\\
1-3 & 7 & 0.3 & 0.94(75) & 5$\pm$0.28 (4.7) & 5.6$\pm^{75}_{5.6}$ & 12$\pm^{34}_{12}$ \\
3-6 & 7  & 0.3& 0.99(46) & 2.51$\pm$0.32 (2.2) &\nodata &\nodata\\
6-9 & 7 &0.3 &  0.79(27) & 1$\pm$0.4 (0.7) (12\%) &\nodata &\nodata\\
\multicolumn{7}{c}{{\bf A2241} (z=0.063, $N_H$=2.05, Ref:[WJF97])}\\
0-1 & 2.1$\pm^{1.6}_{0.4}$ & 0.3 &  0.89(39) & 1.42$\pm$0.13 (2.14) &\nodata &\nodata \\
1-3 & 3.1 &0.3 &  0.83(27) & 0.92$\pm$0.16 (1.17) &\nodata &\nodata\\
3-6 & 3.1 & 0.3 & 0.86(21) & 0.51$\pm$0.26 (0.528) (10\%) &\nodata &\nodata\\
\multicolumn{7}{c}{{\bf A2244} (z=0.097, $N_H$=1.9, Ref:[WJF97])}\\
0-1 & 3.4$\pm^{3.1}_{1.1}$ &0.3 & 1.0(20) & 7.2$\pm$0.5 (7.46) &\nodata &\nodata\\
1-3 & 4$\pm^{4.9}_{1.5}$ & 0.3  & 0.79(56) & 8.3$\pm$0.57 (7.8) &\nodata &\nodata\\
3-6 & 7 & 0.3 &  0.46(24) & 3.56$\pm$0.53 (3.6) &\nodata &\nodata\\
6-9 & 7 & 0.3 & 2.4(14) & 1.14$\pm$0.55  (1.39) (14\%) &\nodata &\nodata\\
\multicolumn{7}{c}{{\bf A2255} (z=0.08, $N_H$=2.51, Ref:[W00,I02])} \\
0-1 & 8 & 0.3 &  0.46(33) & 0.74$\pm$0.08 (0.79) &\nodata &\nodata \\
1-3 & 7 & 0.3 &  1.29(106) & 3.8$\pm$0.18 (3.38) &  21$\pm^{14}_{13}$ & 5.3$\pm^{6}_{5}$\\
3-6 & 7 &0.3 & 0.77(119) & 5.4$\pm$0.26 (4.5) & 19$\pm^{17}_{15.5}$ & 7$\pm^{3.2}_{3}$\\
6-9 & 7&0.3 &  0.84(99) & 2.77$\pm$0.26 (2.59) &\nodata &\nodata\\
9-12 & 6 & 0.3  & 1.36(65) & 2.1$\pm$0.29 (1.32) (19\%) &\nodata &\nodata\\
12-15 & 5 & 0.3  & 0.84(54) & 1.18$\pm$0.31 (0.8) (10\%) &\nodata &\nodata\\
\multicolumn{7}{c}{{\bf A2256} (z=0.06, $N_H$=4.55, Ref:[MSFV98,B91])}\\
0-3 & 7 & 0.3 & 1.05(147) & 5.5$\pm$0.197 (5.28) &\nodata &\nodata \\
3-6 & 7 & 0.3 & 1.13(149) & 6.74$\pm$0.24 (6.5) &\nodata &\nodata\\
6-9 & 7 & 0.3 & 1.29(140) & 3.64$\pm$0.22 (3.3) & 1.2$\pm^{16}_{1.2}$ & 4.2$\pm{4.2}$ \\
9-12 &7 & 0.3 & 0.85(125) & 1.52$\pm$0.21 (1.6) (22\%) &\nodata &\nodata\\
\multicolumn{7}{c}{{\bf A2597} (z=0.085, $N_H$=2.2, Ref:[W00,I02])} \\
0-1 & 2.5$\pm^{0.4}_{0.6}$ & 0.3 & 1.12(140) & 13.1$\pm$0.44 (12.92) &\nodata &\nodata\\
1-3 & 4.3$\pm^{4.2}_{1.5}$ & 0.3 & 0.78(65) & 6.8$\pm$0.34 (4.9) & 58$\pm^{29}_{24}$ & 25$\pm{8.7}$\\
3-6 & 4 & 0.3 & 1.07(35) & 1.84$\pm$0.3 (1.93) &\nodata &\nodata \\
6-9 & 4 &0.3 &  1.04(21) & 0.66$\pm$0.33 (0.69) &\nodata &\nodata \\
\multicolumn{7}{c}{{\bf A2670} (z=0.076, $N_H$=2.73, Ref: [W00,HW97])}\\
0-1 & 3.2$\pm^{2.1}_{0.9}$ & 1$\pm^{3.6}_{0.6}$ & 1.3(61) & 1.27$\pm$0.1 (1.5) & &\\
1-3 & 3 & 0.3 & 1.11(98) & 2$\pm$0.15 (2.85) &\nodata &\nodata\\
3-6 & 3 & 0.3 & 1.09(69) & 2$\pm$0.22 (1.51) & 2.7$\pm^{11}_{2.7}$ & 2.2$\pm{2.2}$ \\
6-9 & 3 & 0.3 & 1.42(46) & 0.63$\pm$ 0.26 (0.68) (10\%) &\nodata &\nodata\\
\multicolumn{7}{c}{{\bf A2717} (z=0.049, $N_H$=1.23, Ref:[L97])} \\
0-1 & 1.8$\pm^{0.4}_{0.25}$ & 0.5  & 1.25(35) & 2.81$\pm$0.18 (2.8) &\nodata &\nodata \\
1-3 & 1.8$\pm^{0.6}_{0.3}$ & 0.19$\pm^{0.29}_{0.14}$  & 1.04(49) & 5.2$\pm$0.28 (5.0) &\nodata &\nodata \\
3-6 & 3.1 $\pm^{4.1}_{1.2}$ & 0.3  & 0.83(44) & 3.3$\pm$0.32 (3.3) &\nodata &\nodata \\
6-9 & 2.7$\pm^{7}_{1.1}$ & 0.3  & 1.12(27) & 1.28$\pm$0.30 (1.26) &\nodata &\nodata \\
\multicolumn{7}{c}{{\bf A2744} (z=0.31, $N_H$=1.41, Ref:[E96,A00])} \\
0-1 & 11 &0.2 & 1.17(50) & 2.52$\pm$0.14 (2.21) & 48$\pm^{88}_{48}$ & 36$\pm^{51}_{36}$  \\
1-3 & 11 &0.2 & 1.25(56)& 3.28$\pm$0.18 (2.38) & 290$\pm{107}$ & 170$\pm{60}$\\
3-6 & 11& 0.2 & 0.77(36) & 1.54$\pm$0.23 (0.92) & 160$\pm^{140}_{148}$& 110$\pm^{73}_{99}$\\
6-9 & 11& 0.2 & 0.83(21) & 0.64$\pm$0.26 (0.32) (9\%)&\nodata &\nodata\\
\multicolumn{7}{c}{{\bf A3301} (z=0.054, $N_H$=2.49)} \\
0-1 & 1.7$\pm^{1}_{0.35}$ & 0.3  & 0.93(14) & 0.51$\pm$0.1 (0.66) &\nodata &\nodata\\
1-3 & 11$\pm^{}_{8}$ & 0.3  & 0.9(34) & 1.54$\pm$0.2 (1.38) &\nodata &\nodata\\
3-6 & 7$\pm^{}_{4}$ & 0.3  & 1.3(41) & 2$\pm$0.31 (1.5) & 9.4$\pm^{10}_{8.5}$ & 4.1$\pm^{4.5}_{4.1}$ \\
6-9 & 2.1$\pm^{}_{0.6}$ & 0.3 & 0.77(35) & 0.57$\pm$0.42 (0.95) (6\%) &\nodata &\nodata\\
\multicolumn{7}{c}{{\bf A3558} (z=0.048, $N_H$=4.0, Ref:[MFSV98])} \\
0-1 & 2.8$\pm^{0.7}_{0.5}$ & 0.48$\pm^{0.36}_{0.22}$ & 1.31(117) & 2.02$\pm$0.3 (1.69) & \nodata &\nodata\\
1-3 & 3.3$\pm^{0.6}_{0.5}$ &  0.26$\pm^{0.16}_{0.15}$ & 1.08(142) & 4.45$\pm$0.14 (4.18) & 0.76$\pm^{9.2}_{0.76}$ & 2.8$\pm{2.8}$ \\
3-6 & 5 & 0.3 & 1.19(150) & 6.86$\pm$0.2 (6)&  15$\pm^{11}_{6}$ & 13$\pm^{5.5}_{4.5}$ \\
6-9 & 5& 0.3  & 1.15(146) & 5.1$\pm$0.23 (3.9) & 28$\pm^{9}_{7}$ & 19$\pm{6.3}$\\
9-12 &5 & 0.3 & 0.98(139) & 2.5$\pm$0.21 (2.1) (22\%) &\nodata &\nodata\\
\multicolumn{7}{c}{{\bf A3560} (z=0.04, $N_H$=4.7, Ref:[W00])}\\
0-3 & 2.1$\pm^{1.3}_{0.6}$ & 0.27$\pm^{0.62}_{0.23}$ & 1.15(39) & 1.23$\pm$0.23 (1.36) & \nodata& \nodata\\
3-6 & 4.1$\pm_{2}$ & 0.3 &  1.25(45) & 1.98$\pm$0.34 (1.29) & 0.8$\pm^{15}_{0.8}$ & 3.6$\pm^{10.5}_{3.6}$ \\
6-12 & 2$\pm^{1.4}_{0.5}$ & 0.4$\pm^{1.4}_{0.3}$  & 1.35 (54) & 2.1$\pm$0.6 (1.4) (14\%) &\nodata &\nodata\\
\multicolumn{7}{c}{{\bf A3562} (z=0.04, $N_H$=4.01, Ref:[I02])} \\
0-1 & 2.4$\pm^{0.8}_{0.4}$ & 0.6  & 1.01(130) & 1.12$\pm$0.088 (1.05) &\nodata &\nodata\\
1-3 & 2.7$\pm^{0.7}_{0.65}$ & 0.6 & 1.02(142) & 2.9$\pm$0.16 (2.5) & 0.4$\pm^{3.8}_{0.4}$ & 2.4$\pm^{3.5}_{2.4}$\\
3-6 & 4.8$\pm^{2.3}_{0.75}$ & 0.4 & 0.82(131) & 3.38$\pm$0.23 (2.7) & 5.2$\pm^{8.2}_{5.2}$ & 4$\pm^{5.2}_{4}$\\
6-12 & 3.6$\pm^{1.1}_{0.8}$ & 0.4 & 1.32(133) & 3.5$\pm$0.4 (3.15) (15\%) &\nodata &\nodata\\
\multicolumn{7}{c}{{\bf A3571} (z=0.04, $N_H$=4.4, Ref:[MSFV98])}\\
0-2 & 5.3$\pm^{5.2}_{1.7}$ & 0.5 &  0.81(118) & 9.03$\pm$0.42 (7.3) & 1$\pm^{9}_{1}$ & 2.6$\pm^{3.8}_{2.6}$ \\
2-4 & 7 & 0.3 & 0.98(129) & 8.65$\pm$0.46 (7.94) & 1.3$\pm^{8.3}_{1.3}$ & 4.5$\pm^{6.7}_{4.5}$ \\
4-6 & 7&0.3 &  1.24(112) & 6.7$\pm$0.45 (5.9)& 1.3$\pm^{23}_{1.3}$ & 13$\pm^{18}_{13}$ \\
6-8 & 7& 0.3&  1.29(76) & 3.9$\pm$0.38 (2.9) & 0.7$\pm^{20.5}_{0.7}$ & 7.4$\pm^{11}_{7.4}$ \\
8-12 &7 &0.3 &  0.92(94) & 4.2$\pm$0.54 (3.27) & 0.25$\pm^{5.2}_{0.25}$ & 3.4$\pm^{15}_{3.4}$\\
12-16 & 5 &0.3 & 1.21(88) & 2.7$\pm$0.62 (1.93) (16\%) &\nodata &\nodata\\
\multicolumn{7}{c}{{\bf A4059} (z=0.046, $N_H$=1.06, Ref: [MFSV98])}\\
0-1 & 2.3$\pm^{0.9}_{0.4}$ & 0.5 &  1.39(64) & 9$\pm$0.41 (10.46) &\nodata &\nodata \\
1-3 & 3.9$\pm^{2}_{1.2}$ & 0.7$\pm^{1.5}_{0.5}$ & 0.86(95) & 16.56$\pm$0.58 (15.16) &\nodata &\nodata \\
3-6 & 5 & 0.3 &  0.71(78) & 11.2$\pm$0.56 (11.64) &\nodata &\nodata \\
6-9 & 4 & 0.3 &  0.97(46) & 4.9$\pm$0.5 (5.9) &\nodata &\nodata \\
9-12 & 4 & 0.3 &  1.05(31) & 0.25$\pm$0.52 (0.28) (19\%) &\nodata &\nodata \\
\multicolumn{7}{c}{{\bf Coma (A1656)} (z=0.023, $N_H$=0.9, Ref: [I02])}\\
0-3 & 8.2 & 0.2 &  1.19(150) & 44.5$\pm$0.48 (36.37) & 11$\pm^{1}_{1.1}$ & 4.6$\pm^{0.56}_{0.3}$\\
3-6 & 8.2 &0.2   & 1.23(150) & 92.9$\pm$0.7 (76.9) & 22 $\pm{1.6}$ & 9.1$\pm{0.7}$\\
6-9 & 8.2& 0.2   & 1.56(150) & 99.1$\pm$0.75 (81.1) & 26$\pm{1.4}$& 10$\pm^{0.86}_{1.1}$\\
9-12 &8.2 & 0.2  & 1.12(150) & 84.8$\pm$0.73 (67.8) & 25$\pm^{1.7}_{2}$ & 10$\pm^{0.7}_{0.5}$\\
12-18 &8.2 & 0.2 & 1.2 (150) & 123$\pm$1 (90.7) & 47$\pm^{2.3}_{2.7}$ & 21$\pm^{0.9}_{0.5}$\\
\multicolumn{7}{c}{{\bf Fornax (AS373)} (z=0.0064, $N_H$=1.5, Ref:[J97])} \\
0-1 & 0.91$\pm^{0.03}_{0.02}$ & 0.47$\pm^{0.07}_{0.05}$ & 1.37(148) &3.16$\pm$0.08 (3.7) &\nodata &\nodata \\
1-2 & 1.28$\pm^{0.04}_{0.07}$ & 0.89$\pm^{0.37}_{0.2}$ &  1.23(149) & 2$\pm$0.073 (1.4) & 0.026$\pm^{0.004}_{0.006}$ & 0.021$\pm^{0.005}_{0.004}$\\
2-3 & 1.38$\pm^{0.04}_{0.04}$ & 1.18$\pm^{0.7}_{0.18}$ &  0.99(149) & 1.7$\pm$0.07 (1.17) & 0.016$\pm^{0.008}_{0.004}$ & 0.017$\pm^{0.0035}_{0.004}$ \\
3-6 & 1.57$\pm^{0.09}_{0.09}$ & 0.98$\pm^{0.28}_{0.18}$ & 0.77(140) & 5.7$\pm$0.15 (4.4) & 0.053$\pm^{0.01}_{0.016}$ & 0.034$\pm{0.007}$\\
6-9 & 1.51$\pm^{0.1}_{0.09}$ & 0.61$\pm^{0.15}_{0.13}$ &  0.92(142) & 5.7$\pm$0.18 (5.0) & 0.027$\pm{0.0016}$ & 0.018$\pm{0.01}$\\
9-12 & 1.5$\pm^{0.08}_{0.07}$ & 0.73$\pm^{0.17}_{0.16}$ & 1.35(142) & 4.97$\pm$0.2 (4.92) &\nodata &\nodata\\
12-15& 1.4$\pm^{0.03}_{0.03}$ & 0.87$\pm^{0.13}_{0.19}$ &  0.94(142) & 0.47$\pm$0.23 (0.465) (23\%) &\nodata &\nodata\\
15-18 & 1.35$\pm^{0.04}_{0.04}$ & 0.71$\pm^{0.17}_{0.16}$ & 0.90(143)& 0.42$\pm$0.25 (0.51) (18\%) &\nodata &\nodata \\
\multicolumn{7}{c}{{\bf Hercules (A2151)} (z=0.037, $N_H$=3.2, Ref:[HS96])} \\
0-1 L & 1.7$\pm^{0.24}_{0.19}$ & 0.5 &  0.96(53) & 1.48$\pm$0.12 (1.5) &\nodata &\nodata \\
1-3 L & 2.4$\pm^{0.9}_{0.5}$ & 0.7 $\pm^{0.8}_{0.35}$  & 0.99(81) & 3.1$\pm$0.2 (2.33) &\nodata &\nodata \\
0-1 R & 1.2$\pm^{0.2}_{0.2}$ & 0.24$\pm{0.7}_{0.2}$ &  0.92(12) & 0.35$\pm$0.073 (0.35) &\nodata &\nodata \\
1-3 R & 1.28$\pm^{0.32}_{0.14}$ & 0.16$\pm^{0.16}_{0.08}$  & 1.02(40) & 0.836$\pm$0.15 (0.82) &\nodata &\nodata \\
\multicolumn{7}{c}{{\bf Virgo} (z=0.0043, $N_H$=1.8)} \\
0-3 & 1.77$\pm^{0.03}_{0.03}$ & 1.14$\pm0.1$ &  1.16(149) & 132$\pm$1.2 (95.6) & 1.8$\pm{0.2}$ & 1.2$\pm{0.075}$\\
3-6 & 2.42$\pm^{0.29}_{0.16}$ & 1.06$\pm^{0.17}_{0.15}$  & 1.1(149) & 110.7$\pm$1.1 (83.6) & 1.6$\pm{0.14}$ & 0.9$\pm{0.08}$\\
6-9 & 2.9$\pm^{0.4}_{0.2}$ & 0.88$\pm^{0.2}_{0.16}$ &  0.94(149) & 98.35$\pm$1.1 (72.2) & 1.8$\pm^{0.14}_{0.16}$  & 0.93$\pm{0.08}$\\
9-12 & 3.6$\pm^{0.7}_{0.4}$ & 0.79$\pm^{0.21}_{0.2}$  & 1.24(149) & 76$\pm$1 (56) & 1.5$\pm^{0.17}_{0.15}$ & 0.7$\pm^{0.05}_{0.06}$\\
12-15 & 3.0$\pm^{0.5}_{0.34} $ & 0.59$\pm^{0.19}_{0.16}$ & 1.0(147) & 62.3$\pm$1 (47.5) & 1.2$\pm^{0.17}_{0.15}$ & 0.54$\pm^{0.06}_{0.07}$\\
15-18 & 3.9$\pm^{1}_{0.7}$ & 1.0$\pm^{0.75}_{0.3}$ &  1.17(149) & 43.2$\pm$1 (32.9) & 0.82$\pm^{0.11}_{0.12}$ & 0.41$\pm^{0.06}_{0.04}$\\
\enddata
\tablecomments{C-band fluxes are shown with their 1-$\sigma$ statistical uncertainties; 
in parentheses, the C-band percent flux above
background for outermost cluster regions, where no percentage is shown
C-band flux exceeds 25\% of C-band background. kT and A errors
are 90 \% confidence intervals,
where no errors are reported, the parameter was fixed and literature reference for T and A
of each cluster is reported. 
 $L_{NT}$ is un-absorbed luminosity in 0.2-0.4 keV
band of the excess emission, when modeled with a power-law model of photon index $\alpha=1.75$,
errors are 90\% confidence on 2 free parameters (power-law normalizations constant
and normalization constant of hot ICM MEKAL model).
$L_{therm}$ is un-absorbed luminosity in 0.2-0.4 keV
band of the excess emission, when modeled with a second MEKAL model of KT=80 eV, A=0.3,
errors are 90\% confidence on 1 free parameter (normalization of second MEKAL model).
Model luminosities have been redshift corrected (K correction), H$_0$=75 km s$^{-1}$ Mpc$^{-1}$.\\
References: \\
 E96 -  Ebeling et al. 96;
 W00 - White 2000; 
 L01 - Lima-Neto, Pislar and Bagchi 2001 (A85);
% M98 - Markevitch et al. 1998;
 R81 - Reichert et al. 1981 (A133);
 BH96 - Briel and Henry 1996 (A1795);
 MFSV98 - Markevitch, Forman, Sarazin and Vikhlinin 1998; 
 HB96 - Henry and Briel 1996 (A2142); 
 MVFS99 - Markevitch et al 1999 (A2199);
 SSJ98 - Siddiqui et al. 1998 (A2199);
 NB99 - Neumann and Boheringer 1999 (A2218);
 WJF97 - White, Jones and Forman 1997; 
 B91 - Briel et al. 1991 (A2256); 
 HW97 - Hobbs and Willmore 1997 (A2670);
 L97 - Liang et al. 1997 (A2717); 
 A00 - Allen 2000; 
 HS96 - Huang and Sarazin 1996 (Hercules); 
% B95 - Briel et al. 1995 (Hercules); 
 J97 - Jones et al. 1997 (Fornax); 
 I02 - Ikebe et al. 2002.} 
\end{deluxetable}

\section{Results}
Results of the spatially resolved C-band emission of all clusters are reported in
Table 3, and 
the average fractional excesses $\eta$ for all clusters are summarized in
Fig. 1. 
Statistically significant detection of
soft excess emission throughout the whole cluster was achieved for A85, A133, A1314, 
A1795, A2218,  A2597, A2744,
A3558, A3562, A3571, Coma, Fornax and Virgo.
In particular, large statistical significance is displayed by those
sources with the deepest observations (Coma, Virgo, Fornax, A3558, A1795).
On the other hand, our available PSPC data appear to rule out excess emission
throughout the entire cluster extent
only in a few sources (A1045, A665, A2241).
The distribution of the soft excess emission/absorption with cluster radius
varies within the sample, and it is shown in detail for each cluster in Fig. 2.
The large number of clusters analyzed in this paper (38) 
warrants however an investigation of the general properties of the sample
as a whole.
The distribution of the soft excess emission is spatially unresolved,
at the resolution of PSPC, in most clusters where excess is found with
large statistical significance (e.g., A1795, Fornax, Coma, Virgo, A2744);
in only two cases the soft excess emission seems to be
localized (A85 and A2256,  see section 6 for details). We therefore focus
on the soft X-ray properties of the azimuthally averaged annuli of all
clusters.

Clusters span an interval of redshifts from z=0.0043
(Virgo) to z=0.308 (A2744); we accordingly plotted the composite
distribution of the C-band fractional excesses $\eta$ of all
cluster regions as function of physical
distance from the cluster center (in Mpc, Fig. 3; a Hubble
constant of $H_0= 75$ km s$^{-1}$ Mpc$^{-1}$ is assumed hereafter).
Only regions for which C-band signals  exceed 25\% of the background
are analyzed, and each region is reported
in the following plots (Figs. 3,5,6,7 and 8) as a data point at its mean radius;
horizontal error bars are omitted for clarity.
When all radii are considered, we found that 91 datapoint-regions lay
above the 0\% excess line, and 38 below it. A closer look at Fig. 3 however
reveals that, although amidst large statistical uncertainties, a
break point in the distribution occurs in the 150-200 kpc
region.
For radii $\leq$ 170 kpc, the distribution reveals a large
scatter, and no evidence of a bias towards
positive $\eta$. Quantitatively, 36 points lay above the no-excess
line, and 27 below it. For radii larger than 170 kpc, all regions
(with the exception of one) are either statistically consistent with 
no excess, or significantly in excess. In numbers, 55 points
exceed $\eta=0$, and 11 have $\eta < 0$.
Fig. 4 summarizes this behavior, whereby the average soft excess
for all cluster regions is $\sim$ 9\% (Fig. 4, top). Central cluster regions,
however, have negligible net average soft excess emission (Fig. 4, center),
while outer cluster regions have an average soft excess of 13\% (Fig. 4, bottom).
A similar result is borne out by Fig. 5, where all data points
are plotted as function of their respective X-ray core radii (see
Table 1). The breakpoint is now located at $\sim$ 0.5-0.6 core radii;
above it, the large majority of data points represent excess
emission.

We also offer an at-a-glance summary of the $\eta$ distribution as
function of detector radius in Fig. 6, primarily for the investigation
of any possible systematic detector effects. 
The excess emission
is confined to detector radii $\geq$ 2 arcmin and, to within the
sparse sampling of our ensemble, rather uniformly distributed
across the
2-18 arcmin region. 
It is also evident, from the individual plots of Fig. 2,
that soft excess regions are detected at different detector positions
and with different radial trends for different clusters, e.g., A1795, A2744, Coma
or A1367~\footnote{All of these observations share the same boresight position
in PSPC detector coordinates.}.
It is therefore possible to rule out the
possibility of detector non-uniformities  and systematic miscalibrations 
as the cause of excess
C-band counts. 
It is also possible to investigate any  correlation of soft excess
emission with $N_H$, given that our sample spans the 0.8--5 $\times 10^{20}$ cm$^{-2}$ range.
Fig. 7 shows no evidence of a correlation of
positive $\eta $ regions with $N_H$. Obviously, at low $N_H$, many of the
detections have larger statistical significance, as more photons
penetrate the Galaxy. However, if the
cross-sections for Galactic absorption were
systematically off,
we would expect a larger fraction of positive $\eta$
regions at one end and a smaller fraction at the opposite end of
the $\eta$-$N_H$ plot. Neither feature is however present in Fig.
7, indicating that cross-section errors are unlikely the
cause of the C-band excess counts.

In Fig. 8 we also plot the composite $\eta$ distribution of
peripheral cluster regions where the C-band  emission
drops below 25\% of the PSPC background (the dotted diamonds
of Fig. 2), and therefore the most sensitive to background
subtraction. All data points are consistent with no excess, indicating
that the employed background subtraction method does not give
rise to systematic under- or over-subtraction of signal.

We summarize the general properties of the C-band emission from
this sample of galaxy clusters in the following points:\\
1. The excess emission co-exists (in a sample sense)
 with intrinsic excess absorption
in the central regions of clusters (radii $\leq$ 150-200 kpc).
This is indicated by the large scatter of the high
signal-to-noise data points in this
radial interval, with a mean value that is approximately
consistent with $\eta$=0. Central soft X-ray absorption is detected
with S/N $\geq$ 3 in 
$\sim$ 20\% of the sources (A665, A2199, A2241, A2670, A4059, Fornax, A1991,
see Fig. 2)\\
2. The soft excess is clearly present outside the central 0.5-0.6
X-ray core radii, or, alternatively, at radii $\geq$ 170 kpc. This
therefore implies that, when the distribution at all physical radii is
considered, excess emission is preferentially found in the outer
regions.\\
3. The soft excess phenomenon is widespread. Not only $\eta >$0
regions are the typical feature  of Figs. 1 through 7, but excess is detected with
large statistical significance (e.g., S/N $\geq$ 3-4) preferentially
in the deepest observations, as expected of any genuine celestial phenomenon
(see A1795, A85, Coma, Virgo, Fornax, A3571 and A3558; A2199 is the only cluster for which 
a deep PSPC
observation did not reveal soft excess with large
S/N). \\
4. The soft excess emission seldom exceeds 40-50\% of the hot ICM
contribution. It is evident that clusters like S\'{e}rsic 159-03
(Bonamente, Lieu and Mittaz 2001d), where the soft excess reaches a prominence of 50-100\%
in PSPC C-band, are
exceptional. Few cases where comparison with   {\it EUVE}/DS data in the $\sim$ 65-190 eV band is
available (Coma, Virgo, A2199, A1795; Lieu et al. 1996a,1996b; Mittaz, Lieu and Lockman 1998; 
Lieu, Bonamente and Mittaz 1999) suggest that the
soft excess phenomenon may be intrinsically very {\it soft}, e.g.
more suitably revealed with large $\eta$ in the extreme-ultraviolet band.

\section{Interpretation}
The analysis reveals that the customary one-temperature hot ICM
model fails to describe accurately the C-band emission
of many clusters in this sample.
 Positive and negative residuals $\eta$ are
present in the majority of sources, and constitute a common trait
in the composite plots of the whole sample.
Central cluster regions (i.e., $\leq$ 150-200 kpc from the center) 
show evidence of intrinsic absorption in a fraction of the sources, and of
excess emission in others (e.g., A133, Coma and Virgo).
For radii $\geq$ 200 kpc, the 1-T model typically falls short of the detected
emission, revealing the soft excess emission.

{(a) {\it C-band central absorption}} \\
Intrinsic C-band absorption can be attributed to cold (T $\leq
10^4$K) gas in the center of galaxy clusters. There the hot ICM has
radiative cooling times which are often significantly shorter than
the cluster's  age, and the hot gas is expected to cool
to low temperatures (Fabian  1988).
There is also evidence, from PSPC image analysis (Bonamente, Lieu and Mittaz 2001a),
 that cold absorbing gas may co-exists with the hot gas even in
 the absence of the excess
 absorption signature. Presence of cold, absorbing gas could then
 be responsible for excess C-band absorption in some clusters,
 and for the lack or lesser C-band excess emission in other
 clusters where the excess is evident at larger radii (e.g.,
 A1795, Fornax etc.).

{(b) {\it Thermal and non-thermal modeling of the excess emission}} \\
The nature of the excess component has been studied in detail for
those sources for which the data quality allows a detailed
modeling. The excess emission of Coma (Lieu et al. 1996b),
Virgo (Lieu et al. 1996a; Bonamente, Lieu and Mittaz
2001b), A2199 (Lieu, Bonamente and Mittaz 1999),
 A1795 (Bonamente, Lieu and Mittaz 2001b), A3558 and A3571 (Bonamente et al. 2001c;
and of S\'{e}rsic 159-03; Bonamente, Lieu and Mittaz 2001d,  not included in this sample) indicate that
the emission may be due to either a second phase of the ICM at
lower temperature (T $\sim  10^6$ K, the thermal `warm' gas scenario),
or to an Inverse-Compton emission
of cluster relativistic electrons that scatter microwave
background photons (MWB, non-thermal model).
Here we model the excess emission of those cluster regions
where excess emission is detected with S/N $\geq$ 1 $\sigma$, in order to
provide estimates/upper limits to the soft X-ray luminosity
of the excess component ($L_{NT}$ and $L_{therm}$ in Tables 3, 5, 6 and 7).
Employed emission models are: \\
\indent 1. an optically thin plasma emission code (MEKAL) of kT=80 eV and A=0.3 solar (thermal model);\\
\indent 2. a power-law  with photon index $\alpha$ = 1.75 
(non-thermal model). This emission model is suggestive of Inverse Compton emission
due to scattering of MWB photons
with a population of relativistic electrons with
differential number distribution n(E) $\propto$ E$^{-p}$, $p$=2.5 (e.g.,
Lieu et al. 1999).\\
The two models indicate that the soft excess emission
has a 0.2-0.4 keV luminosity which ranges between $10^{40}$ erg s$^{-1}$ and 
10$^{43}$ erg s$^{-1}$,
the latter figure applies to the strong soft excesses of Coma and A2744.
In summary, either model successfully explains the
data, but faces serious theoretical
challenges.
Sub-megakelvin gases cool radiatively in a
short time by free-free and line emission.
Our estimates require a warm gas density
of $\sim 10^{-3}$ cm$^{-3}$ or more (if the gas is substantially clumped)
  and T$\sim 10^6$ K in order to explain typical soft excess
fluxes (Bonamente, Lieu and Mittaz 2001b,2001d). The free-free cooling time of such gas,
\begin{equation}
\rm t_{cool}\simeq 6 \times 10^{9} ( \frac{T}{10^6 K}) ^{\frac{1}{2}}
                    ( \frac{n}{10^{-3} cm^{-3}})^{-1} \; \rm years
\end{equation}
is by itself already
shorter than the Hubble time, and line cooling for metal
abundances of A$\sim$ 0.3 solar further reduces the cooling time by a
factor of about 5, or even more if the gas is cooler than $10^6$ K
(Landini and Monsignori-Fossi, 1990).
A mechanism of continuous
injection/reheating would therefore be necessary to sustain the emission for cosmological
periods of time, (e.g., Fabian 1997) as its frequent occurrence is unlikely
to be explained as a transitory phenomenon.
On the other hand, the non-thermal model requires large amounts
of cosmic-ray electrons, which may surpass the energy content of
the hot phase (see, e.g., Bonamente, Lieu and Mittaz 2001d).

In order to obtain more precise estimates of the `warm' gas masses and the non-thermal electron
pressure implied by the soft excess it is necessary to know the
precise temperature of the putative `warm' gas, or alternatively
the slope of the non-thermal power-law that best fits the soft excess.
This information is unavailable for most of the
clusters analyzed here, due to limited S/N. The case of S\'{e}rsic 159-03,
where the fractional excess reaches $\eta =0.5-1$ in PSPC C-band
(Bonamente, Lieu and Mittaz 2001d), can however be used as an upper limit for
all clusters investigated in this sample. Its soft excess, when
modeled as thermal emission, implied that an excess of $\sim 40-50$\%
by mass (with respect to calculated hot ICM masses)
resides as a warm phase of the ICM. Alternatively,
relativistic electrons are present in the ICM above pressure equipartition
with the hot ICM by a factor of few.
After a downward revision to match the present $\eta \sim 0.1-0.5$ fractional
excesses of this sample,
the two scenarios imply that in a large
fraction of galaxy clusters ($\sim$ 50\%) there is either an extra 10-40\%
of mass at sub-Virial temperatures, or a widespread population of relativistic
electrons near pressure equipartition with the hot gas.

{(c) {\it Radial distribution of the excess and non-thermal models}}\\
Superior spectral resolving power is required
to distinguish  the signatures of `warm' gas and  of a non-thermal power-law
from X-ray spectroscopy, marginally available even to the
new generation X-ray instruments (XMM/EPIC and RGS, Chandra/LETG
and ACIS). Alternatively, other wavelengths are also employed in the
search of, e.g., thermal OVI lines from metal-rich sub-Virial gases in cluster,
although yet unsuccessfully (e.g., Dixon et al. 2001);
these results will be rendered obsolete if the putative warm gas is primordial.
However, the soft excess radial distribution of clusters in this sample
provides by itself additional clues toward understanding  the nature of the
phenomenon.
The preferential distribution of soft excess emission at large
cluster radii is naturally explained in the context of non-thermal
Inverse Compton (IC) radiation. The distribution of the relativistic electrons as
function of cluster radius $r$
(n$_e (r)$)
may follow the gas distribution,
while the MWB photons are
uniformly distributed  across the sky. With the hot ICM gas density distribution n$(r)$  declining
with radius, the hot ICM bremsstrahlung emission falls with
 radius like n$^2 (r)$, while the IC emission only like n$_e
(r)$=n$(r)$. Alternatively, if thermal and non thermal electrons
coexist in an adiabatic environment, the different ratio of
specific heats $\gamma$ (5/3 for the hot ICM and 4/3 for the
relativistic plasma) predicts a larger fraction of relativistic electrons
at large cluster radii than at small radii with respect to the thermal gas, going further in the
direction of explaining the presence of  soft excess at large radii (e.g., Bonamente 2000).
It is however possible, as discussed early in the section, that
central excess emission is `masked' by intrinsic absorption, and
that the increasing radial trend is not a genuine feature of the emitting
mechanism.

{(d) {\it Cluster mergers and the soft excess}} \\
A few clusters in this sample, notably the two rich clusters
A2142 and A2256,  show evidence of recent mergers through
asymmetric distributions of X-ray emission.
In A2256 there is evidence that soft excess emission is associated
with the merging sub-cluster, also of higher X-ray brightness,
 located to the east of
the main brightness peak. In the case of
A2142, we find no evidence of soft excess emission associated with
the bright south-western irregular quadrant.
In both cases the excess X-ray brightness associated with the
putative merging event can be due to extra thermal radiation from
the merging group;
but it is also easily explained with a non-thermal power-law,
encompassing 0.2-2 keV luminosities of same order as the thermal luminosity in those regions,
see Tables 6 and 7.
Following the latter interpretation, a cluster in the early merger phases may
develop significant X-ray excess emission through acceleration
of relativistic electrons at strong merger shocks (e.g., Fujita and Sarazin 2001).
A few Giga-years after the primary acceleration, the merging subcluster tends to virialize and to assume a more isotropic
X-ray brightness distribution; at the same time, relativistic
electrons undergo severe interim losses (e.g., Lieu et al. 1999;
Ip and Axford 1985) which preferentially leave relic electrons
in the $\Gamma \sim$ 200-400 Lorentz factor range, distributed over larger volumes
than the original acceleration sites.
The relic electrons emit IC radiation precisely
in the EUV and soft X-ray band, as the energy of emitted photons
relates to the electron Lorentz
factor via the relationship
\begin{equation}
\Gamma=(\frac{3 \epsilon}{4 \epsilon_{MWB}})^\frac{1}{2}
\end{equation}
where $\epsilon$ is emitted energy, and $\epsilon_{MWB}$ is the average
energy of a MWB photon. The soft excess
may therefore be the effect of a merging event occurred a few Giga-years ago,
as also shown in detail by Lieu et al. (1999) for the Coma cluster.

{(e) {\it Warm-hot intergalactic filaments and the soft excess}}\\
A new way of interpretation is provided by recent hydrodynamic simulations
of formation and evolution of large-scale structures. General agreement among
most researchers is that a large fraction of the current epoch's baryons
reside mainly in filamentary structures of  a warm-hot intergalactic medium (WHIM),
extending for several mega-parsecs with T=$10^5-10^7$ K and overdensities of
$\delta \sim 5-200$ relative to the critical density (e.g., Cen and Ostriker 1999).
%($\rho_{cr} = 1.0 \times 10^{-29}$ g cm$^{-3}$ for $H_0$=75 km s$^{-1}$ Mpc$^{-1}$).
These filaments may in fact
converge toward galaxy clusters (e.g., Cen et al. 2001, Dave' et al. 2001).
The details of the soft X-ray emission from such filamentary structures
heavily depend on their temperature, density and metal abundances, which
at the moment can only be predicted within order-of-magnitude ranges.
However, a simple estimate indicates that a
 1 Mpc length of a $2 \times 10^{-4}$ cm$^{-3}$, T=$3 \times 10^6$ K filament
directed towards the observer,
yields a free-free surface brightness at 0.25 keV which amounts to $\sim$10\% of that of
a typical cluster line of sight encountering a 1 Mpc-long column of hot
ICM gas at T=$10^8$  K and density $10^{-3}$ cm$^{-3}$.
In fact,
the critical density  (if $\Omega_0=1$ and H$_0$=75 km s$^{-1}$ Mpc$^{-1}$) is
\begin{equation}
\rm \rho_{cr}=\frac{3 H_0^2}{8 \pi G}=1 \times 10^{-29}\; g\; cm^{-3}= 6 \times 10^{-6}\; H \;atoms \;cm^{-3}.
\end{equation}
 An overdensity of $\delta=30$ implies
a H atoms density of $n_{f}$=1.8  $\times 10^{-4}$ cm$^{-3}$ (if $\Omega_b \sim \Omega_M$).
On the other hand, typical hot ICM gas densities are of the order of
$n_{ICM}$=$10^{-3}-10^{-4}$ H atoms cm$^{-3}$.
The optically thin free-free emission per unit volume in  a given band near frequency $\nu$ is
proportional to
\begin{equation}
E \propto  n^2 \times T^{-1/2} e^{-\frac{h \nu}{k T}}
\end{equation}
where $kT$ is the temperature of emitting gas, and $h \nu \sim 0.25$ keV applies to the C-band.
As an example, we consider the emission from warm filaments at kT=240 eV (T $\sim 3 \times 10^6$ K)
and $n_{f}$=1.8  $\times 10^{-4}$ cm$^{-3}$ ($\delta$ = 30),
and emission from an hot ICM medium at kT=8.2 keV
and $n_{ICM}$=$10^{-3}$ cm$^{-3}$, reminiscent of the Coma cluster.
The ratio of filament-to-ICM soft X-ray detected surface brightness is
\begin{equation}
r= \frac{E_{f} \times D_{f}}{E_{ICM} \times D_{ICM}} = \frac{D_{f} \times n_{f}^2 T_f^{-1/2} e^{-\frac{h \nu}{k T_{f}}}}
{D_{ICM} \times  n_{ICM}^2 T_{ICM}^{-1/2} e^{-\frac{h \nu}{k T_{ICM}}}}
\end{equation}
where $D_{f}$ is projected length (in cm) of the filament along the line of sight,
and similarly $D_{ICM}$ is the length of the cluster's H column observed, assumed
to have a mean density of $n_{ICM}$.
For the current example, if we assume $D_{f}=D_{ICM}$ (say, 1 or 2 Mpc),
the above formula yield r=0.09, or free-free emission from
the warm filament is 9\% of the hot ICM emission.
Clearly $r$ is a sensitive function
of the filament's temperature, of the gas temperatures and of
the extent of the ICM and filament gas columns, which may vary significantly
from case to case.
Moreover, warm filaments at T $\sim 10^6$ K  may have substantial additional
soft X-ray emission
in the form of lines,
 which will
increase the ratio $r$ by a factor of several (e.g., Landini and Monsignori-Fossi 1990).
% (see Appendix for details).

It is therefore possible that the soft excess emission of many clusters 
is in fact the result of the detection of WHIM filaments projecting themselves onto the clusters.
If the filaments indeed converge toward galaxy clusters, it is possible to
surmise that lines of sight toward these sources have a higher probability of
intercepting filaments with relatively larger overdensities, and
therefore
filaments are more easily detectable through their projection onto the
central 1-2 Mpc of galaxy clusters. 
Different filament geometries and alignments with
respect to our line of sight may be responsible for the different radial trends of the
excess emission reported in this paper.

{(f) {\it Cluster galaxies and the soft excess}}\\
We also entertained the possibility of
the soft excess originating from a large number of
unresolved, X-ray emitting cluster galaxies, as 
galaxies of all morphological types are spatially extended sources of
X-ray emission (e.g., Fabbiano 1989). 
Recent high-resolution Chandra observations (Blanton, Sarazin
and Irwin 2001;
Sarazin, Irwin and Bregman  2001) indicate that X-ray emission from early type
(E and S0) galaxies can be resolved into a diffuse `warm' gas halo at kT $\sim$ 0.5 keV,
and a contribution from low mass X-ray binaries (LMXB) and other stellar systems.
Late type galaxies, on the other hand, normally lack a diffuse halo, but may host
X-ray emitting LMXBs, like in the case of our own Milky Way (e.g., Liu et al. 2000,2001,
Fabbiano 1989).
The diffuse galactic halo emission is significantly softer than that of the hot ICM;
on the other hand, the softeness of the LMXB radiation, typically composed
of a kT$\sim$0.25 keV black-body component and a harder $\alpha \sim 1.2$ power-law
(e.g., Blanton, Sarazin and Irwin 2001), depends on the relative ratio of the two components.

While X-ray bright galaxies can be easily resolved and excluded from the analysis
of the diffuse cluster soft X-ray emission, X-ray faint galaxies
cannot be identified at the resolution of the present study.
X-ray faint galaxies have soft X-ray luminosities of the order of $\sim 10^{38}$
erg s$^{-1}$ (e.g., Fabbiano, Kim and Trinchieri 1992); this figure requires that
several thousand such galaxies  be present to explain typical
soft excess luminosities of order $10^{41}-10^{42}$ erg s$^{-1}$ (Table 3).

Optical studies indicate that the projected galaxy number density distribution in clusters is often well described by a King profile (Dressler 1978):
\begin{equation}
\sigma_{gal}(r)=\sigma_0 \left( 1+ \left( \frac{r}{r_{c,gal}} \right)^2 \right)^{-1}
\end{equation}
where $r$ is cluster projected radius and $r_{c,gal}$ is the galaxy density core radius; its
behavior at large radii is $\propto r^{-2}$. The X-ray surface brightness
of the hot ICM gas, on the other hand, is normally well described by
\begin{equation}
I_{gas}(r) = I_0 \left( 1+ \left( \frac{r}{r_{c,gas}} \right)^2 \right)^{-3 \beta + \frac{1}{2}}
\end{equation}
where $r_{c,gas}$ is the ICM core radius; since $\beta$=2/3 in most clusters,
 $I_{gas}(r) \propto r^{-3}$ at large radii. The two formulas
indicate that the X-ray emission from individual galaxies can be more extended than the hot
ICM radiation, and it can therefore
potentially explain the larger fractional soft excess emission found at large
distances from the cluster's center (Fig. 4).

In summary, the excess soft X-ray emission can in principle be due to a very large
number of unresolved, X-ray faint galaxies, through their stellar and diffuse
halo emission. It is however necessary to point out two possible shortcomings of this
interpretation.
Firstly, the very bright excess emission of, e.g., A1795, Coma or A2744, is unlikely 
explained by this scenario, due to the very large number of required faint galaxies ($\geq$ 10,000);
as an example, the Coma cluster has a population of only about 500 galaxies within B magnitude M$_B$=16
(Doi et al. 1995).
Secondly, it is found that early type galaxies are more commonly found in the
central regions of clusters, while the outskirts are
relatively more densely populated by late type galaxies (Doi et al. 1995), generally lacking the
very soft X-ray halo. Excess emission at large radial distances would therefore need to originate
predominantly by stellar emission (LMXB) in late type galaxies alone.

{(g) {\it Additional considerations}}\\
It is also important to remark that even clusters with similar
properties, e.g., Virgo and Fornax, both nearby and poor, or A1795
and A2199, both with bright cD galaxy at the center, central cooler gas (CF, Table 1) and
relaxed X-ray morphology, show different soft excess
morphology and luminosity (see Table 3). It is therefore possible that
soft excess photons are produced by more than one mechanism.

\section{Comments on individual clusters}
{\bf A21} There is no evidence of either excess absorption
or emission.

{\bf A85} Bagchi, Pislar and Lima-Neto (1998) noted two regions of X-ray
enhancement in SAX data, and excess X-ray emission is also evident
in our PSPC data. One is a {\it very steep spectrum radio source} (VSSRS, located near
R.A.=00 41 34.7, Dec.=-9 22 0.2, $\sim$ 5 arcmin offset from the cluster's center
and size $\sim$ 1.75 arcmin radius),
the other region (RADIO2, near R.A.=0 41 48.3, Dec.=-9 28 0.2, $\sim$ 8 arcmin south
of the cluster's center and size 3 arcmin radius) is also
related to a radio source.
We investigated these two regions by (a) studying separately the C-band emission
of the two VSSRS and RADIO2 regions, and that of the remainder of
the corresponding annular regions (ann-VSSRS and ann-RADIO2);
(b) fitting the X-ray excess emission, above the
azimuthal average of the annular region, of VSSR and RADIO2
with a non-thermal power-law spectrum, see Table 4.
In case (a), there seems to be C-band excess emission associated with both
regions,  while little excess is evident in the rest of the annulus;
for case (b), the X-ray excess is well fitted by the power-law model, and
no further C-band excess is noted. It seems therefore that A85's soft excess
is localized in the two radio emission regions and of non-thermal origin.
\begin{deluxetable}{lccccc}
\tablecaption{The soft X-ray properties of the VSSRS and RADIO2 regions of A85,
and their azimuthally averaged annuli (A-VSSRS and A-RADIO2)}
\tablehead{ Region & $\chi^2_r$ (dof) & C-band flux (model) & $\alpha$ & EM & $L_{nt}/L_{th}$ \\
    (arcmin)     &                  & $10^{-2}$ c s$^{-1}$ &            &           &                 \\}
%\colhead{Region} & \colhead{$\chi^2_r$ (dof)} & \colhead{C-band flux (model)} & 
%\colhead{$\alpha$} & \colhead{EM excess} & \colhead{$\frac{L_{nt}}{L_{th}}$}}
\startdata
\hline
VSSRS   &   0.65(57) & 1.75$\pm$0.16 (1.43) & \nodata &\nodata & \nodata \\
A-VSSRS &  0.82(151) & 11.8$\pm$0.4 (11.5) & \nodata & \nodata & \nodata\\
RADIO2 &  0.63(78) & 2.64$\pm$0.2 (2.12) & \nodata & \nodata & \nodata\\
A-RADIO2 &  0.82(144) &9.6$\pm$0.48 (8.9) & \nodata&\nodata & \nodata \\
%\hline
%Region &  $\alpha$ & red. $\chi^2$ (dof) & EM excess  & $L_{nt}/L_{th}$   \\
%\hline
%\tablehead{\colhead{Region} & \colhead{red. $\chi^2$ (dof)} & \colhead{C-band flux (model)} &
%\colhead{} & \colhead{}}
VSSRS XE & 0.7(70) & \nodata & 1.54$\pm^{0.21}_{0.15}$ & $\sim$14.2  & $\sim$1.0  \\
RADIO2 XE& 0.64(99) & \nodata &1.5$\pm^{0.11}_{0.14}$ &  $\sim$12.8 & $\sim$2.2  \\
\hline
\enddata
\tablecomments{`EM' indicates by how many times the normalization constant
 of the spectrum (the Emission Measure) exceeds the prediction from the annular average,
`XE' means X-ray excess.
$L_{nt}/L_{th}$ is the ratio of non-thermal to thermal luminosity
in 0.2-2 keV band. Hot ICM is modeled with a kT=7 keV, A=0.3 MEKAL model at all radii,
HI column density is $N_H= 2.5 \times 10^{20}$ cm$^{-2}$.}
\end{deluxetable}

{\bf A133} There is clear evidence of soft excess
emission in this cluster, from the central regions to $\sim$
500 kpc, the limiting radius of the PSPC detection.

{\bf A665} A665 is one of the furthest clusters of the
sample, z=0.182, and relatively high HI column density ($N_H=4.3 \times 10^{20}$
cm$^{-2}$). It shows evidence of central absorption, and no
soft excess emission.

{\bf A1045} No literature references were available for
temperature and abundances of this cluster; PSPC measures a T $\sim$ 3-5 keV
when metal abundances are fixed at A=0.3 solar (Table 3).
There is marginal evidence of central excess
absorption and no evidence of soft excess emission.

{\bf A1068} Only a 2-$\sigma$ evidence of soft excess in the regions surrounding the
cluster core is evident from our observation.

{\bf A1302} The 2.3 ks PSPC observation is of limited
quality, but compatible with soft excess emission of order 10-50\%
 outside the central 100 kpc, inward of which  the C-band radiation is on
par with the hot ICM.

{\bf A1314} The very short PSPC exposure (2 ks)
of this poor but very nearby cluster indicates,
although within large statistical error bars, that
A1314 may host soft excess emission, throughout the central $\sim$ 500 kpc
where cluster emission is detected.

{\bf A1361} The 3.9 ks PSPC exposure
constrains the C-band excess emission of this cluster to
$\sim $ +20$\pm$ 20\% outside the central 100 kpc.

{\bf A1367} A1367 is an X-ray faint and very nearby
galaxy cluster, near the brighter Coma cluster. It has a secondary
peak of the X-ray brightness (Donnelly et al., 1998), which we
study here separately. The secondary peak is located near R.A.=11 44 21.7, Dec.=+19 52 13.2,
approximately 19 arcmin offset from the primary brightness peak.
\begin{deluxetable}{lcccc}
\tablecaption{Soft X-ray properties of the secondary brightness peak of A1367.} 
\tablehead{ Region & $\chi^2_r$ (dof) & C-band flux (model) & $L_{NT}$&$L_{therm}$ \\
     (arcmin)     &                  & $10^{-2}$ c s$^{-1}$ & 10$^{41}$ erg s$^{-1}$ & 10$^{41}$ erg s$^{-1}$ }
%\colhead{Region} & \colhead{red. $\chi^2$ (dof)} & \colhead{C-band fuls (model)} &
%\colhead{ $L_{NT}$ 10$^{41}$ erg s$^{-1}$} & \colhead{$L_{therm}$ 10$^{41}$ erg s$^{-1}$}}
\startdata
% Region &  red. $\chi^2$ (dof) & 20-41 flux (model) & $L_{NT}$ & $L_{therm}$ \\
%        &                      &                    & 10$^{41}$ erg s$^{-1}$ & 10$^{41}$ erg s$^{-1}$ \\
%\hline
0-3 peak2 & 1.04(71) & 2.8$\pm$0.16(2.4) & 0.54$\pm^{1.7}_{0.54}$ & 0.32$\pm{0.24}$\\
3-6 peak2 & 1.05(119) & 5.96$\pm$0.25 (5.36)(55\%) & 1.1$\pm^{0.8}_{0.9}$ & 0.49$\pm^{0.35}_{0.37}$\\
\enddata
\tablecomments{The hot ICM
emission is modeled with a kT=4.2 keV, A=0.2 solar and HI
column density $N_H= 1.7 \times 10^{20}$
cm $^{-2}$ photoabsorbed MEKAL model. See caption of Table 3 for definition of $L_{NT}$
and $L_{therm}$.}
\end{deluxetable}
While azimuthally averaged annuli around the primary peak revealed
no evidence of soft excess emission (see Fig. 1), the secondary
peak (divided into 0-3 and 3-6 arcmin regions from its center) 
seems associated with soft excess emission, see Fig. 9 and Table 5.

{\bf A1413} At z=0.14, this cluster reveals marginal excess
emission between 0.1-1 Mpc, at a level not exceeding $\sim$30\%.

{\bf A1689} This cluster has no clear evidence of soft
excess emission.

{\bf A1795} A1795 is known to have soft excess emission
from {\it EUVE}/DS observations (Bonamente, Lieu and Mittaz 2001b). The {\it EUVE}
excess was more prominent at large radii (see e.g., Fig. 4 of Bonamente, Lieu and Mittaz 
2001b)  while
the PSPC data analyzed here shows soft X-ray excess emission extending from the
central regions to the limits of the C-band detection, at the level of $\eta =$10-20\%.
There is no evidence of C-band excess absorption in this analysis of PSPC
data.

{\bf A1914} This clusters shows a very
similar behavior to A1361, with possible soft excess emission not
to exceed $\sim$ 20-30\%.

{\bf A1991} There is  evidence of  excess
 absorption in the center and marginal soft excess in the outskirts of  this X-ray faint cluster.

{\bf A2029} This X-ray bright and rich
cluster shows evidence (at $\sim 3-4 \sigma$
level) of soft excess emission outside its central 100 kpc, at level of 10\%.

{\bf A2142} A2142 shows a clear enhancement of the X-ray
emission in the south-eastern quadrant (Briel and Henry 1996). We investigated
the properties of the emission in that region, and how it compares with the
emission in the remainder of the azimuthal directions (see Fig. 9 and Table 6).
\begin{deluxetable}{lccccccc}
\tabletypesize{\small}
\tablecaption{Soft X-ray properties of the south-eastern quadrant of A2142 
(3-6-SE and 6-9-SE) and of the remainder of the azimuthal directions 
(ann-3-6 and ann-6-9)}
\tablehead{Region&$\chi^2_r$ (dof)&C-band flux (model) & $L_{NT}$ & $L_{therm}$&$\alpha$&EM&
     $\frac{L_{nt}}{L_{th}}$ \\
   (arcmin)    &             & 10$^{-2}$ c s$^{-1}$ & 10$^{41}$ erg s$^{-1}$ & 10$^{41}$ erg s$^{-1}$ & & &}

%\colhead{Region} & \colhead{red. $\chi^2$ (dof)} & \colhead{C-band flux (model)} &
%\colhead{$L_{NT}$} & \colhead{$L_{therm}$} & \colhead{$\alpha$} & \colhead{EM excess} & \colhead{$L_{nt}/L_{th}$ }}
\startdata
%\hline
% Region &  red. $\chi^2$ (dof) & 20-41 flux (model) &  $L_{NT}$ & $L_{therm}$ \\
%        &                       &                   & 10$^{41}$ erg s$^{-1}$ & 10$^{41}$ erg s$^{-1}$ \\
ann-3-6 & 0.81(151) & 3.1$\pm$0.19 (2.54) & 6.4$\pm^{20}_{6.4}$ & 12$\pm^{9}_{12}$ &\nodata &\nodata &\nodata\\
3-6-SE &  0.69(151) & 2.05$\pm$0.15 (2.16)&\nodata &\nodata &\nodata &\nodata &\nodata\\
ann-6-9 & 0.75(114) & 1.1$\pm$0.18 (1.44) (23\%) &\nodata &\nodata&\nodata &\nodata &\nodata\\
6-9-SE &  0.53(99) & 1.05$\pm$0.14 (1.14) (46\%) &\nodata &\nodata&\nodata &\nodata &\nodata\\
%Region &  $\alpha$ & red. $\chi^2$ (dof) & EM excess  & $L_{nt}/L_{th}$   \\
%\hline
3-6-SE XE &0.73(179)&\nodata&\nodata&\nodata&1.3$\pm^{0.1}_{0.12}$ &2.5 & 1.35\\
6-9-SE XE &0.58(115)&\nodata&\nodata&\nodata&1.27$\pm^{0.18}_{0.2}$ &2.4 & 1.4\\
\enddata
\tablecomments{The X-ray excess emission (`XE') is also modeled
with a non-thermal power-law, see caption of Table 4 for details. The hot ICM
emission is modeld with a kT=9 keV, A=0.3, $N_H=4.2 \times 10^{20}$ cm$^{-2}$ photoabsorbed
MEKAL model for all regions. See caption of Table 3 for definition of $L_{NT}$
and $L_{therm}$.}
\end{deluxetable}
There is marginal evidence of soft excess emission in the central 500 kpc
of this clusters, detected out to a radius of $\sim$ 1 Mpc.
The south-eastern quadrant's X-ray excess (above the azimuthal average)
 can be
successfully modeled with a power-law model,
indicating a possible non-thermal origin of the X-ray excess brightness.

{\bf A2199} A2199 has marginal evidence of soft excess
emission in this deep PSPC observation. {\it EUVE} data (Lieu, Bonamente and
Mittaz 1999; Bonamente, Lieu and Mittaz 2001b) show significant
extreme-ultraviolet excess that reaches an importance of a few times
100\% at large distances from the center. The only possibility to
reconcile the two observations is that the soft excess emission in
this source is mainly confined to energies below $\sim$ 0.2 keV.

{\bf A2218} There is evidence of soft excess emission at
the level of $\sim$ 20\% in the PSPC observation of this very
rich, z=0.171 cluster.

{\bf A2219} The 8.5 ks observation of this z=0.23 cluster
indicate that soft excess emission may be present at the 20-30\%
level in the regions surrounding the cluster's core, although is
presently detected with low statistical significance.

{\bf A2241} There is evidence of excess absorption in the
central $\sim$ 100 kps of this cluster, which is detected only out
to a radius of $\sim 300$ kpc.

{\bf A2244} The short PSPC exposure detects no evidence
of excess C-band emission/absorption.

{\bf A2255} There is evidence of soft excess emission in
the regions immediately surrounding the core and, with marginal
S/N, at the cluster's outskirts.

{\bf A2256} The X-ray emission of this cluster shows an
asymmetry in the western quadrant (approximately 90 degrees wide, 
positioned 20 degrees west of north to 20 degrees south of west, 
e.g., Briel et al. 1991, Sun et al. 2002), presumably associated
with merger activity.
Concentric annuli are accordingly divided in two sectors, `W' being the
western quadrant,  and the
remainder of the azimuthal directions, to further study the C-band emission (Table 7).
The marginal soft excess
emission detected in the whole cluster seems associated only with the
narrow quadrant of asymmetric X-ray distribution, 
where it is detected with large S/N,
see Fig. 9.
\begin{deluxetable}{lccccccc}
\tabletypesize{\small}
\tablecaption{\footnotesize The soft X-ray properties of 
A2256 in western quadrant (W) and in the remainder of the
azimuthal directions}
\tablehead{Region&$\chi^2_r$ (dof)&C-band flux (model) & $L_{NT}$ & $L_{therm}$&$\alpha$&EM&
       $\frac{L_{nt}}{L_{th}}$ \\
   (arcmin)    &             & 10$^{-2}$ c s$^{-1}$ & 10$^{41}$ erg s$^{-1}$ & 10$^{41}$ erg s$^{-1}$ & & &}
%\tablehead{\colhead{Region} & \colhead{red. $\chi^2$ (dof)} & \colhead{C-band flux (model)} &
%\colhead{$L_{NT}$} & \colhead{$L_{therm}$} & \colhead{$\alpha$} & \colhead{EM excess} & \colhead{$L_{nt}/L_{th}$ }}
\startdata
%\hline
% Region &  red. $\chi^2$ (dof) & 20-41 flux (model) & $L_{NT}$ & $L_{therm}$ \\
%        &                       &                   & 10$^{41}$ erg s$^{-1}$ & 10$^{41}$ erg s$^{-1}$ \\
ann-0-3 & 1.1(138) & 3.1$\pm$0.15 (3.29) & \nodata & \nodata & \nodata & \nodata & \nodata \\
ann-3-6 & 1.04(141) & 3.32$\pm$0.18 (3.8) & \nodata& \nodata&  \nodata & \nodata & \nodata  \\
ann-6-9 & 1.07(131) & 2.31$\pm$0.18 (2.39) &\nodata& \nodata & \nodata & \nodata & \nodata \\
ann-9-12& 0.83(110) & 0.83$\pm$0.18 (1.1) (11\%)& \nodata& \nodata& \nodata & \nodata & \nodata  \\
0-6 W & 1.37(136) & 4.7$\pm$0.18 (3.8) & 65$\pm^{11}_{13}$ & 23$\pm^{7.2}_{6.5}$ &\nodata &\nodata &\nodata\\
6-9 W & 0.92(98) & 1.5$\pm$0.12 (1.17) & 22$\pm{10.5}$ & 7.6$\pm^{4.5}_{3.9}$ &\nodata &\nodata &\nodata\\
%Region & $\alpha$ & red. $\chi^2$ (dof) & EM excess & $L_{nt}/L_{th}$  \\
0-6 XE & 1.0(164) & \nodata& \nodata& \nodata&  1.8$\pm^{0.06}_{0.1}$ & 1.5 &  0.7 \\
6-9 XE & 0.86(116) & \nodata& \nodata& \nodata& 1.8$\pm^{1.1}_{0.4}$ & 1.4 & 0.4  \\
\enddata
\tablecomments{See caption of Table 4 for details. The hot ICM emission
is modeled with a kT=7 keV, A=0.3 and $N_H=4.6 \times 10^{20}$ cm$^{-2}$ photoabsorbed
MEKAL model. See caption of Table 3 for definition of $L_{NT}$
and $L_{therm}$.}
\end{deluxetable}
In addition, the excess X-ray emission in W above the azimuthal
average of the remainder of azimuthal directions, of lower X-ray brightness, is 
modeled with 
a non-thermal model.
The excess X-ray component in 0-6 and 6-9 arcmin regions of W is 
also succesfully modeled with a second thermal componenent at 
T $\sim 1.7-1.8$ keV.

{\bf A2597} The soft excess emission is strongly detected
only in the 1-3 arcmin region, while the C-band signal is
consistent with hot ICM prediction in the inner and outermost
regions.

{\bf A2670} Hobbs and Willmore (1997) indicate that this
cluster has evidence of merger activity. It is strongly absorbed
in the central $\sim$ 200 kpc, and only marginal evidence of
excess outwards.

{\bf A2717} There is no evidence of either excess
emission or absorption in this cluster.

{\bf A2744} The most distant cluster of this sample
(z=0.31, Abell richness class 3) has clear evidence of strong soft excess emission in our
13.8 ks PSPC observation, at the level of $\sim$ 50\%.
The cluster is a strong X-ray emitter, and is detected out to a radius of $\sim$ 2
Mpc. Detected soft excess emission implies thermal and non-thermal luminosities of order $10^{43}$
erg s$^{-1}$ in 0.2-0.4 keV band, see Table 3, the highest to date.

{\bf A3301} A3301 has only marginal evidence of soft
excess emission in the regions outwards of $\sim$50 kpc.

{\bf A3558} The strong soft X-ray excess of A3558 was
already reported by Bonamente et al. (2001c); the excess reaches $\sim$
30\% importance with respect to the hot ICM.

{\bf A3560} There is only marginal evidence of excess
emission; the PSPC exposure achieves only limited S/N.

{\bf A3562} The soft excess of this cluster is revealed
with good statistical confidence in the $\sim$ 400 kpc region of
the PSPC detection (see also Bonamente et al. 2001c).

{\bf A3571} A3571 has a $\sim$ 20\% soft excess emission
extending across the entire cluster extent. A pointed
BeppoSAX observation (Bonamente et al. 2001c) reports the 
matching LECS 0.15-0.3 keV excess.

{\bf A4059} There is no evidence of soft X-ray excess, and
only marginally of intrinsic absorption in the center.

{\bf Coma} The Coma cluster had its soft excess emission
discovered originally with {\it EUVE} (Lieu et al. 1996b), and this PSPC
observation confirms the excess emission with very large
statistical significance. The PSPC excess has an importance of $\sim$
20-30\% of the hot ICM throughout the limits
 of our analysis (18 arcmin radius).

{\bf Fornax} This is a poor and very nearby cluster
(z=0.0046). The deep PSPC observation (53 ks) shows
strong soft excess emission that reaches a 50\% importance (see
also Rangarajan et al. 1995), with
a radial trend resembling closely that of S\'{e}rsic 159-03
(Bonamente, Lieu and Mittaz 2001d), although X-ray emission is detected out
to a radial distance of only $\sim$ 100 kps from the cluster's
center.

{\bf Hercules} The Hercules cluster of galaxies (also
known as A2151) has faint X-ray emission. It shows two peaks of the
X-ray intensity (see Bird et al. 1995, and Huang and Sarazin
1996), which we investigated separately in Fig. 2. There is only marginal
evidence of soft excess emission in the brightest of the two
peaks.

{\bf Virgo} Along with Coma, it was one of the
two clusters where the soft excess emission was initially discovered, in
a deep {\it EUVE} observation (Lieu et al. 1996a). The PSPC excess
emission, already reported also in Bonamente, Lieu and Mittaz (2001b), is
strongly detected with $\eta \sim$ 0.3-0.4 throughout the limits
of this investigation (18 arcmin radius). This cluster is at very
low redshift (z=0.0043), similar to that of the Fornax cluster; it shows a
rather flat radial trend of the soft excess, completely different
from that of Fornax over the same interval of radii ($\sim$ 0-100 kpc).

\section{Conclusions}
The PSPC observations analyzed in this paper 
show excess soft X-ray radiation (0.2-0.4 keV) in a large fraction of the sources,
proving that the thermal radiation from the hot ICM is
not solely responsible for the majority of cluster soft X-ray spectra.
The main indication as to the nature of 
the excess emission is provided by the preferential presence at large
cluster radii, suggesting a non-thermal nature of the phenomenon. 
An alternative  explanation would require  a `warm' ICM phase, although
its short cooling time necessitates a source of re-heating. 
The latter requirement may however
be avoided if the soft X-ray excess originates from the emission of
warm, lower density large-scale filamentary structures impinging on the clusters. 

The authors thank Dr. E.M. Murphy for making his data available ahead of publication and
Dr. Osmi Vilhu for helpful discussions. MB and RL thank NASA for support.

\section*{Figure captions}
Figure 1: The fractional excess $\eta$ for each cluster, see Eq. 1,
averaged over all
cluster regions where C-band emission exceeds 25\% of the background, see Table 1. 

Figure 2: The radial distribution of the fractional soft 
X-ray excess emission/absorption $\eta$ for each cluster, as function of distance from
the cluster's center (bottom axis: Arcmin; top axis: Mpc).
Vertical semidiameters are 1-$\sigma$ errors which include statistical and model uncertanties (see text).
Dotted diamonds are regions where C-band emission is below 25\% of the C-band background; where 
no dotted diamonds are reported, either the PSPC boresight position did not allow to extend the analysis to larger
radii (A1367, Coma and Virgo),
or the peculiar cluster morphology impeded it (Hercules).

Figure 3: Composite $\eta$ distribution of all cluster regions as function of radial
distance. Each data point represents an annular region at its mean radius, horizontal 
error bars are omitted for clarity. Errors are 1-$\sigma$ uncertanties.

Figure 4: The cumulative distribution of $\eta$ 
for all regions (top; $< \eta >$=0.088, variance $\sigma_{\eta}$=0.167), 
only regions at a mean distance $\leq$ 0.17 Mpc
from the cluster center (middle; $< \eta >$=0.024, $\sigma_{\eta}$=0.057) 
and regions at mean distance $\geq$ 0.17 Mpc (bottom; $< \eta >$=0.131, $\sigma_{\eta}$=0.202).

Figure 5:  Composite $\eta$ distribution of all cluster regions as function of X-ray core radius, see Table 1.

Figure 6:  Composite $\eta$ distribution of all cluster regions as function of detector radial distance
from boresight.

Figure 7:  Composite $\eta$ distribution of all cluster regions as function of Galactic  HI column  density, see Table 1. 

Figure 8:  Composite $\eta$ distribution of peripheral cluster regions where C-band counts are $\leq$ 25\% of C-band background.

Figure 9: Left: the $\eta$ distribution in concentric annuli of A1367 around the primary
brightness peak (solid diamonds) and around the secondary peak (dashed diamonds).
Middle: the $\eta$ distribution of A2142 in concentric annuli (solid diamonds, as 
in Fig. 2), and separately in south-western quadrant (dashed diamonds) and remainder of
azimuthal directions (dot-dashed diamonds).
Right: the $\eta$ distribution of A2256 (solid diamonds, as in Fig. 2)
and separately in eastern quadrant (dot-dashed diamonds) and remainder of azimuthal directions (dashed diamonds).

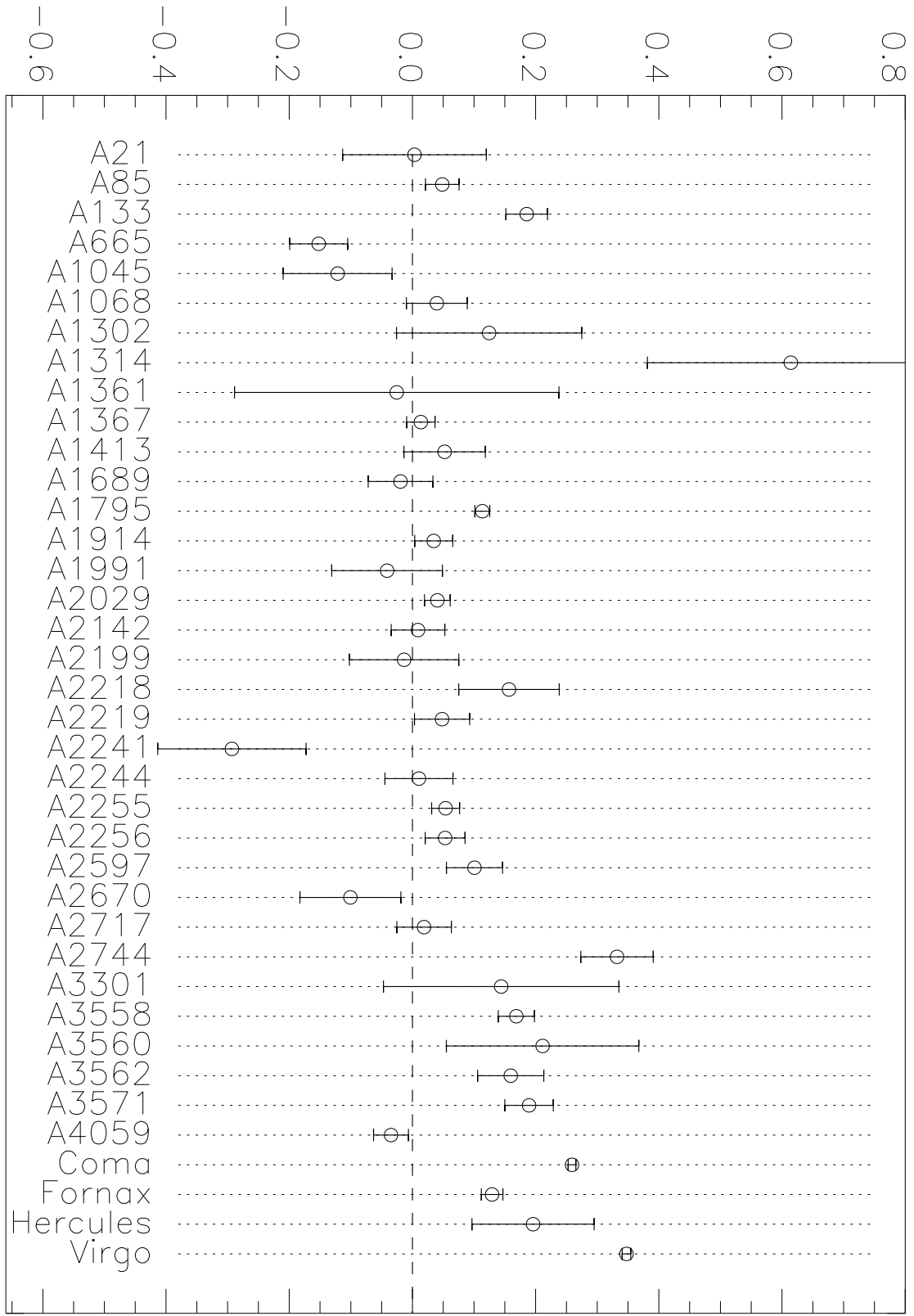
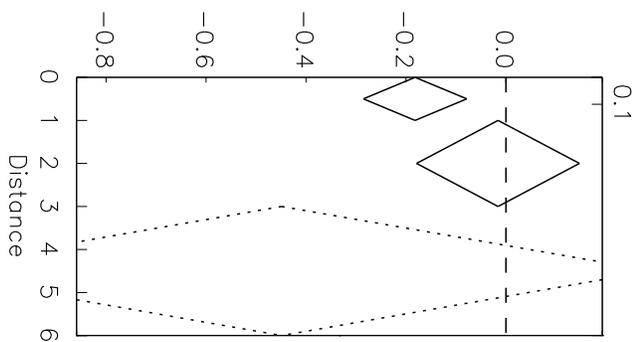
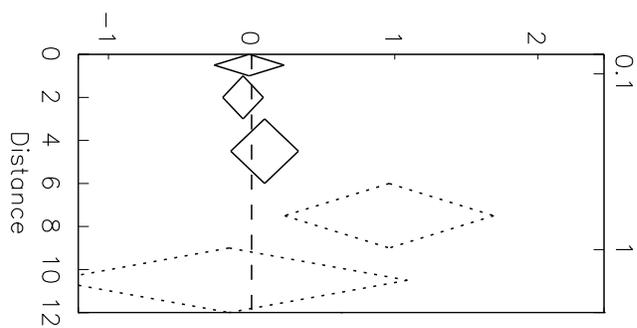
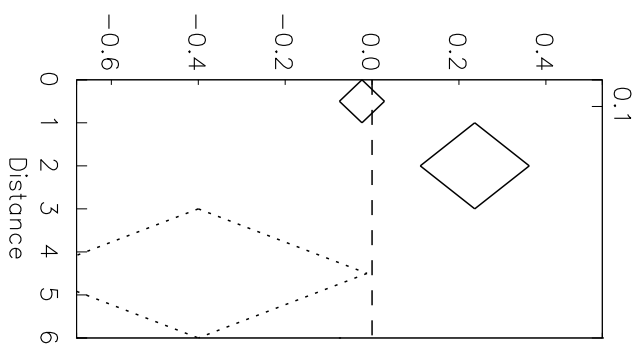
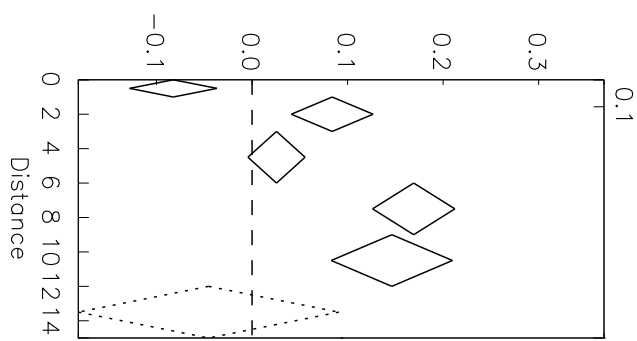
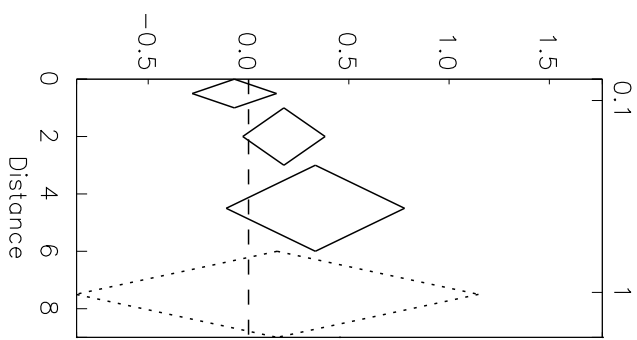
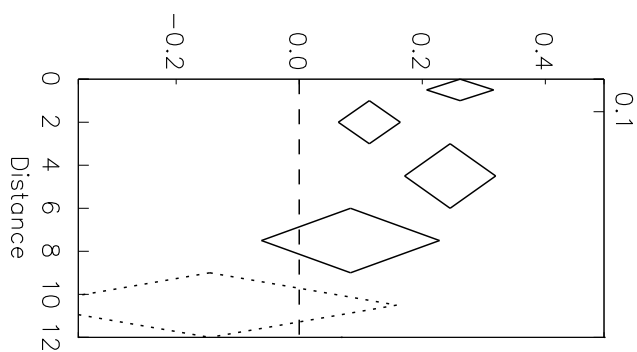
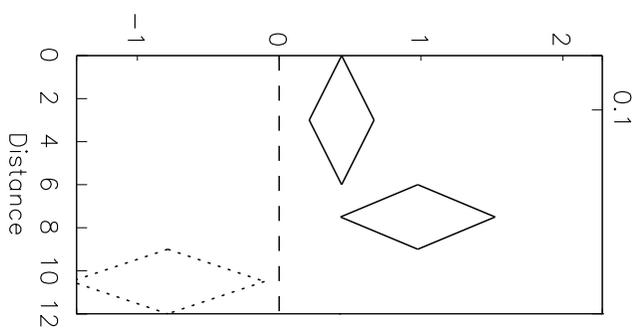
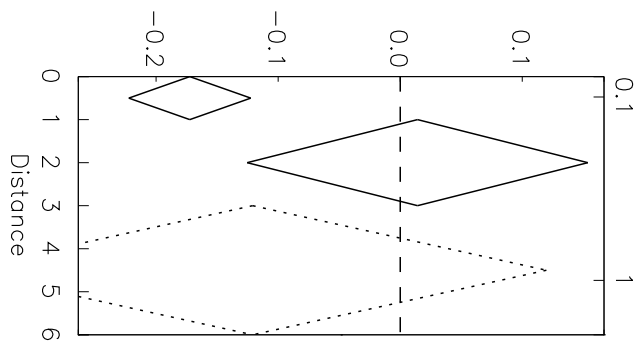
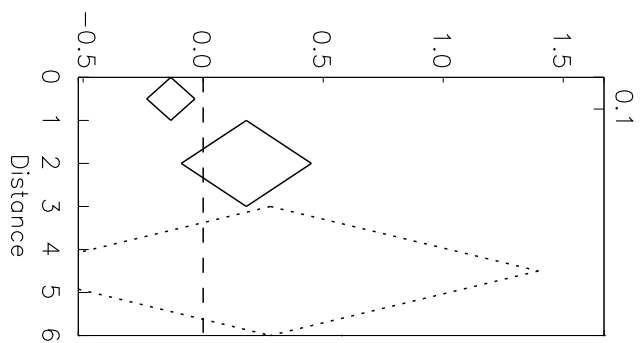
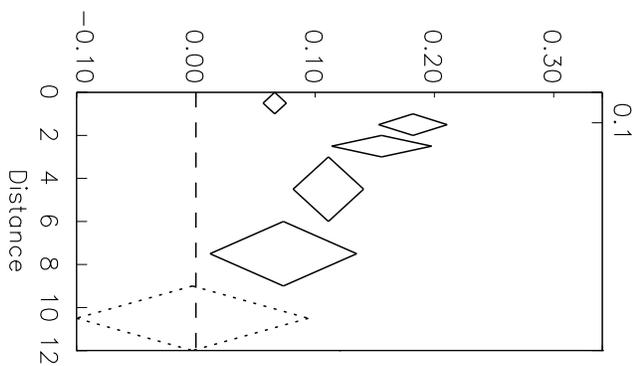
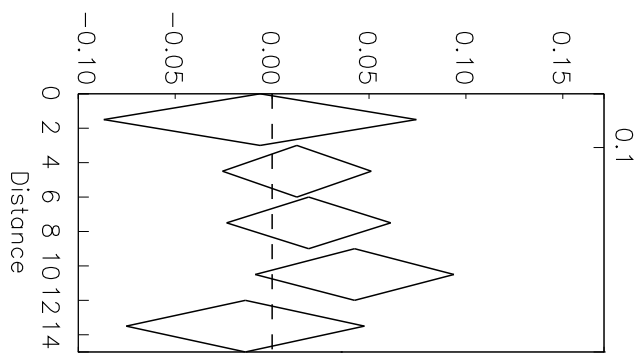
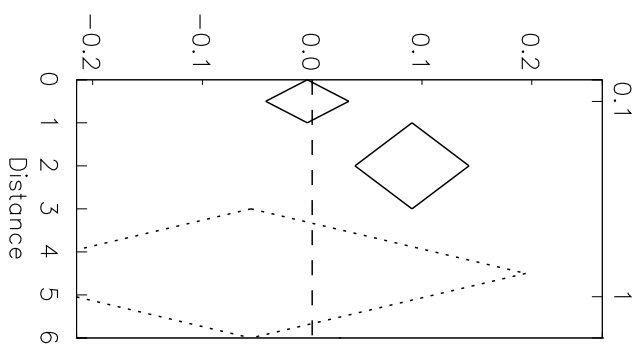
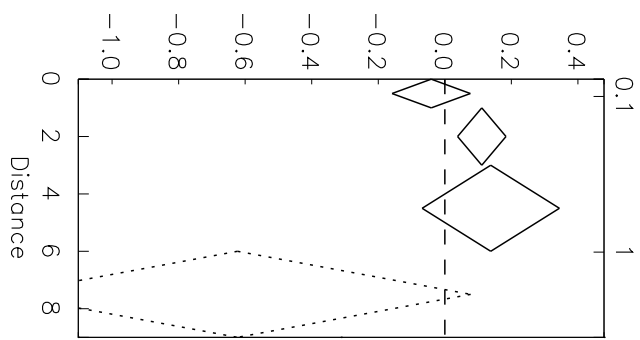
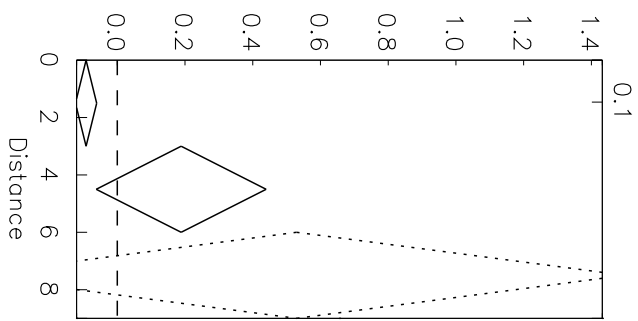
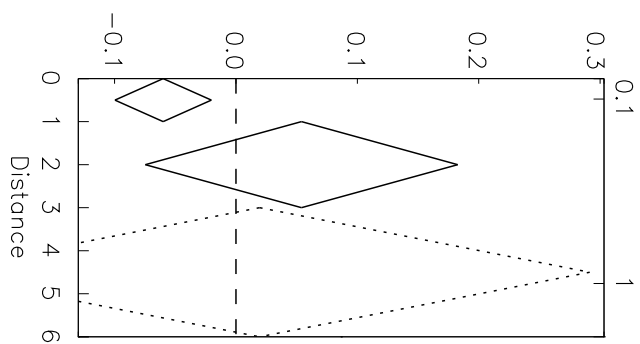
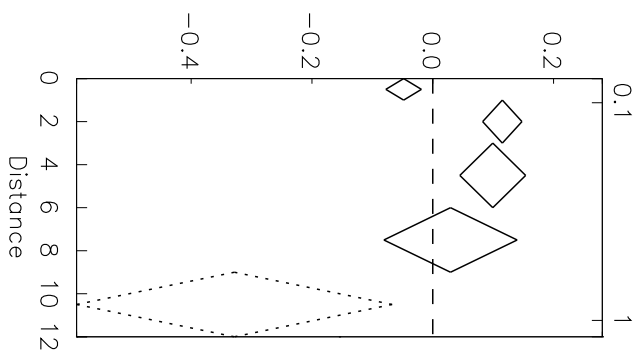
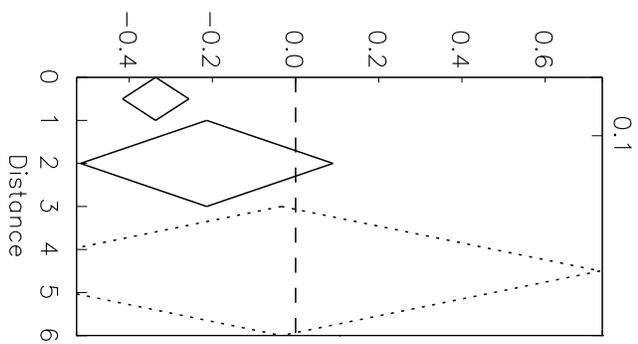
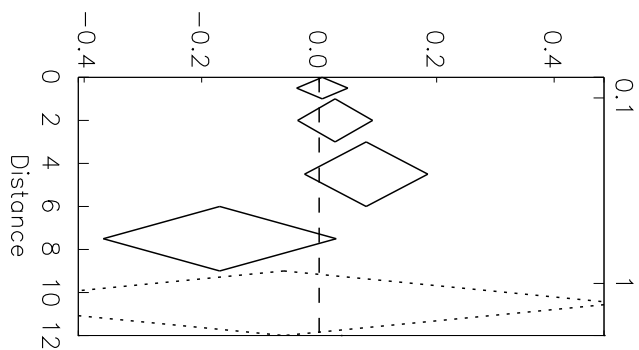
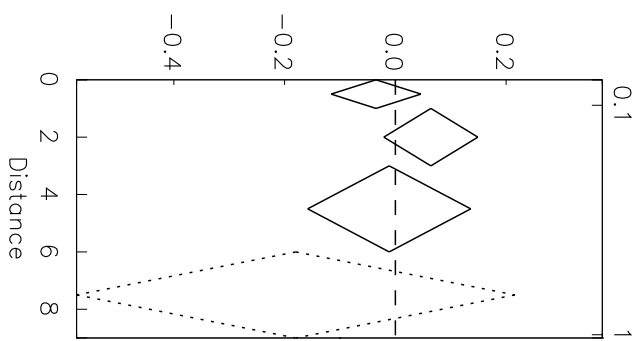
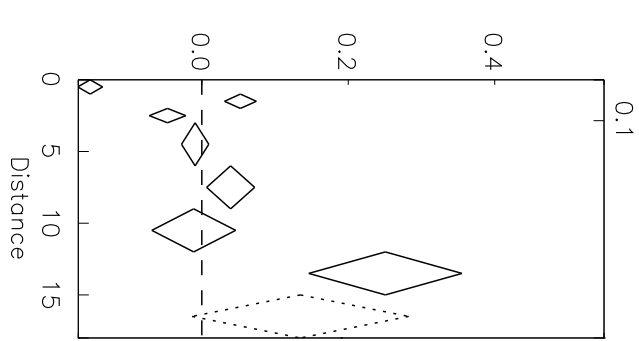
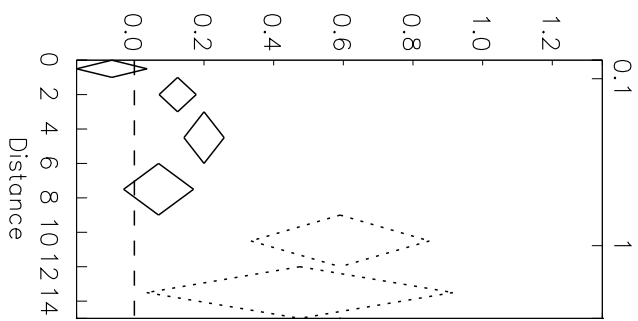
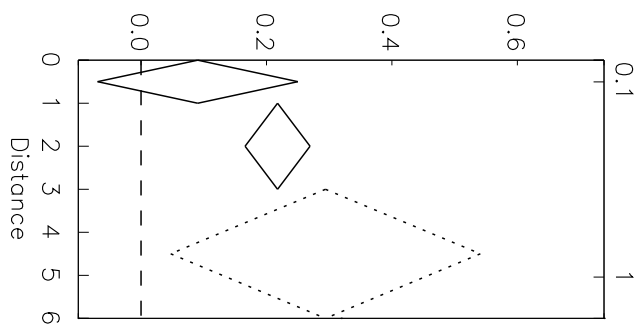
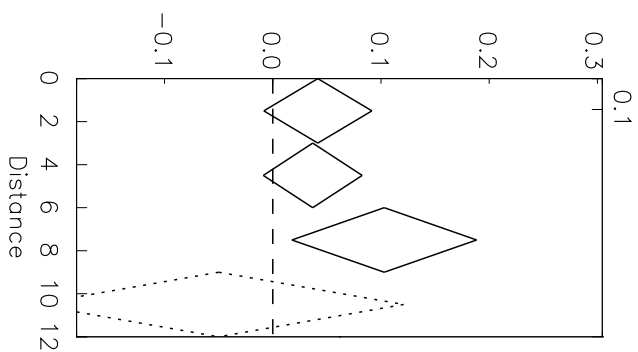
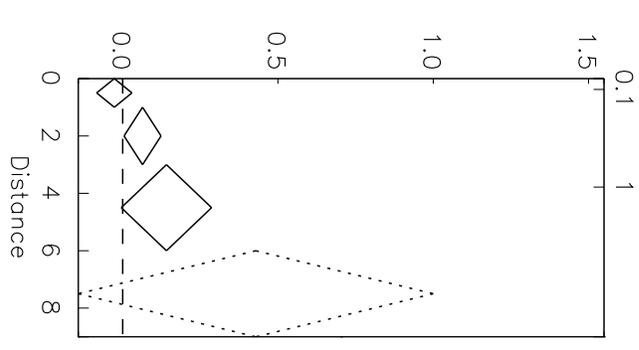
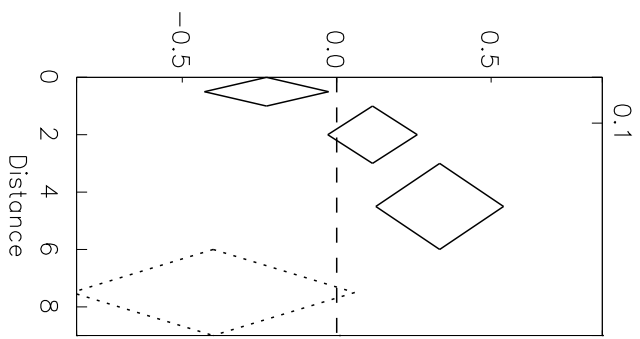
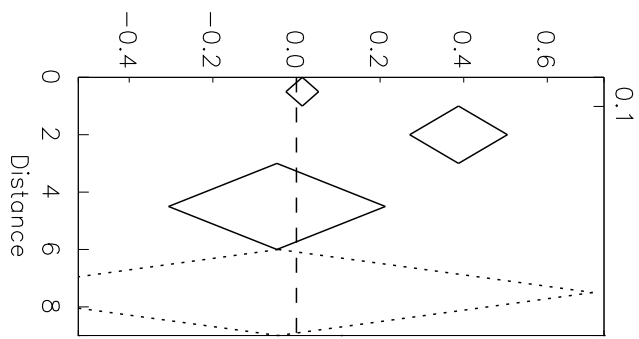
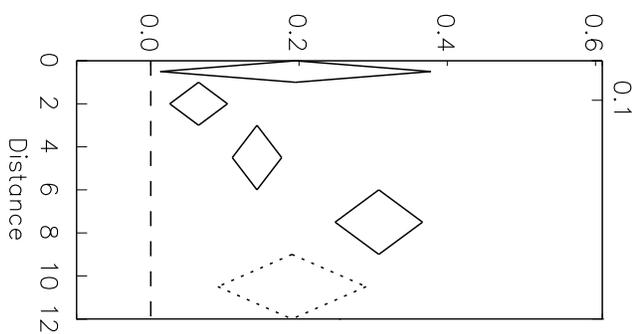
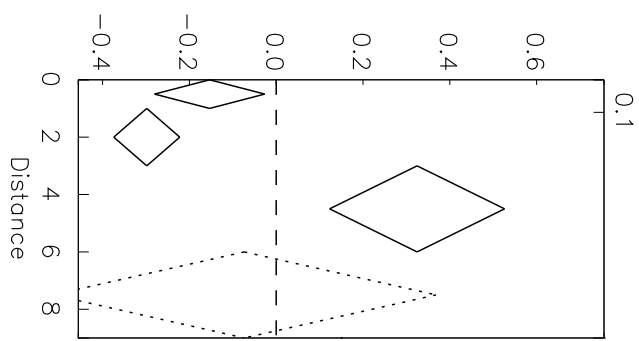
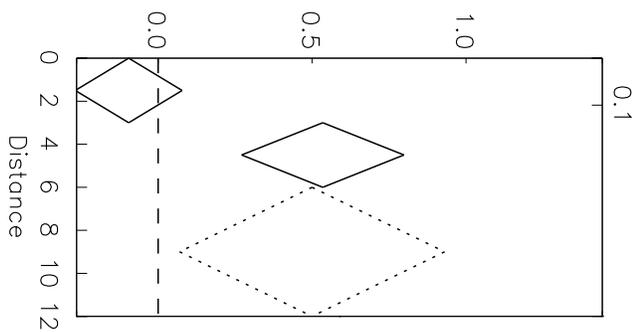
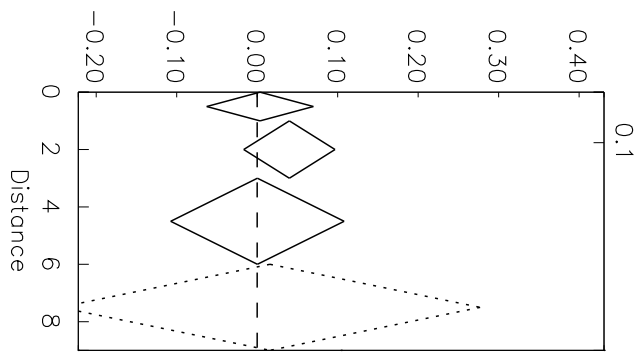
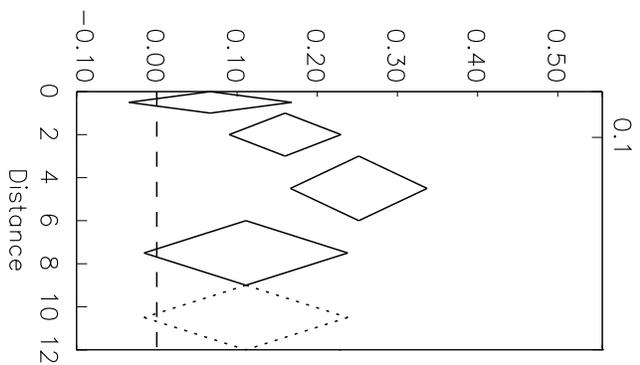
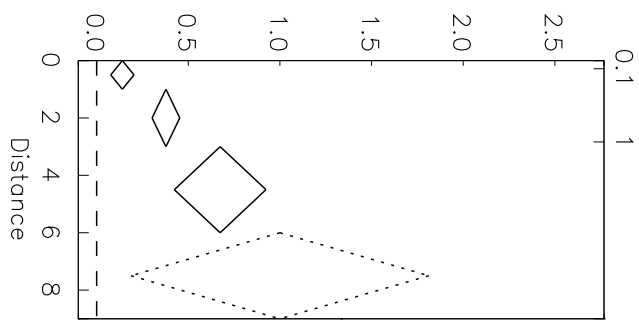
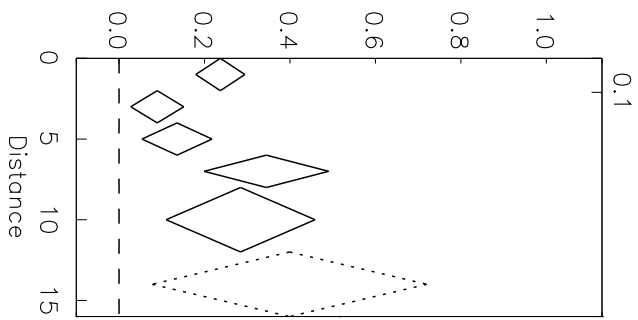
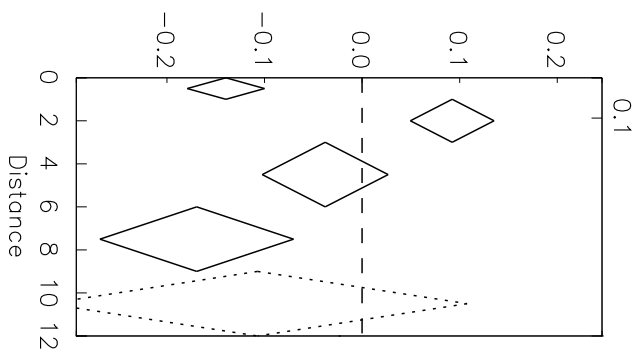
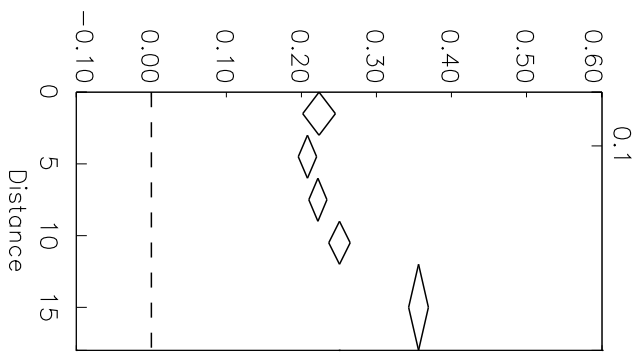
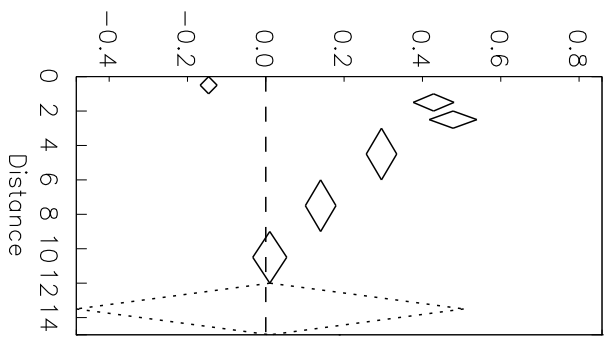
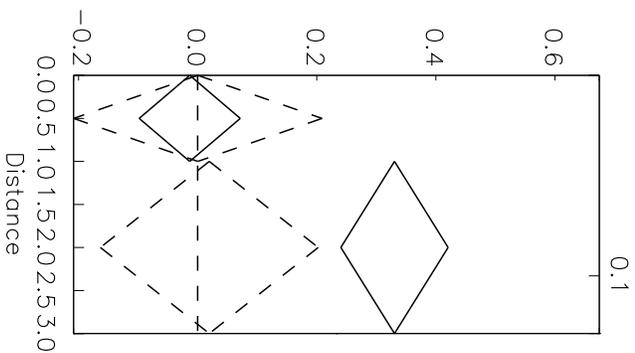
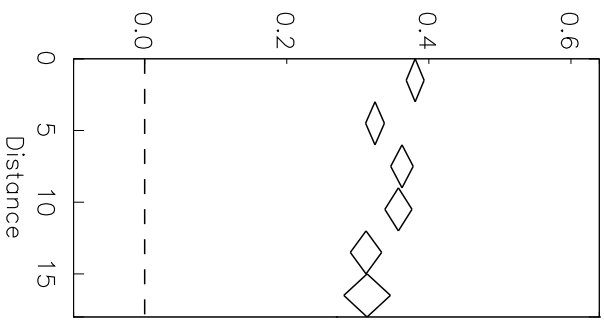
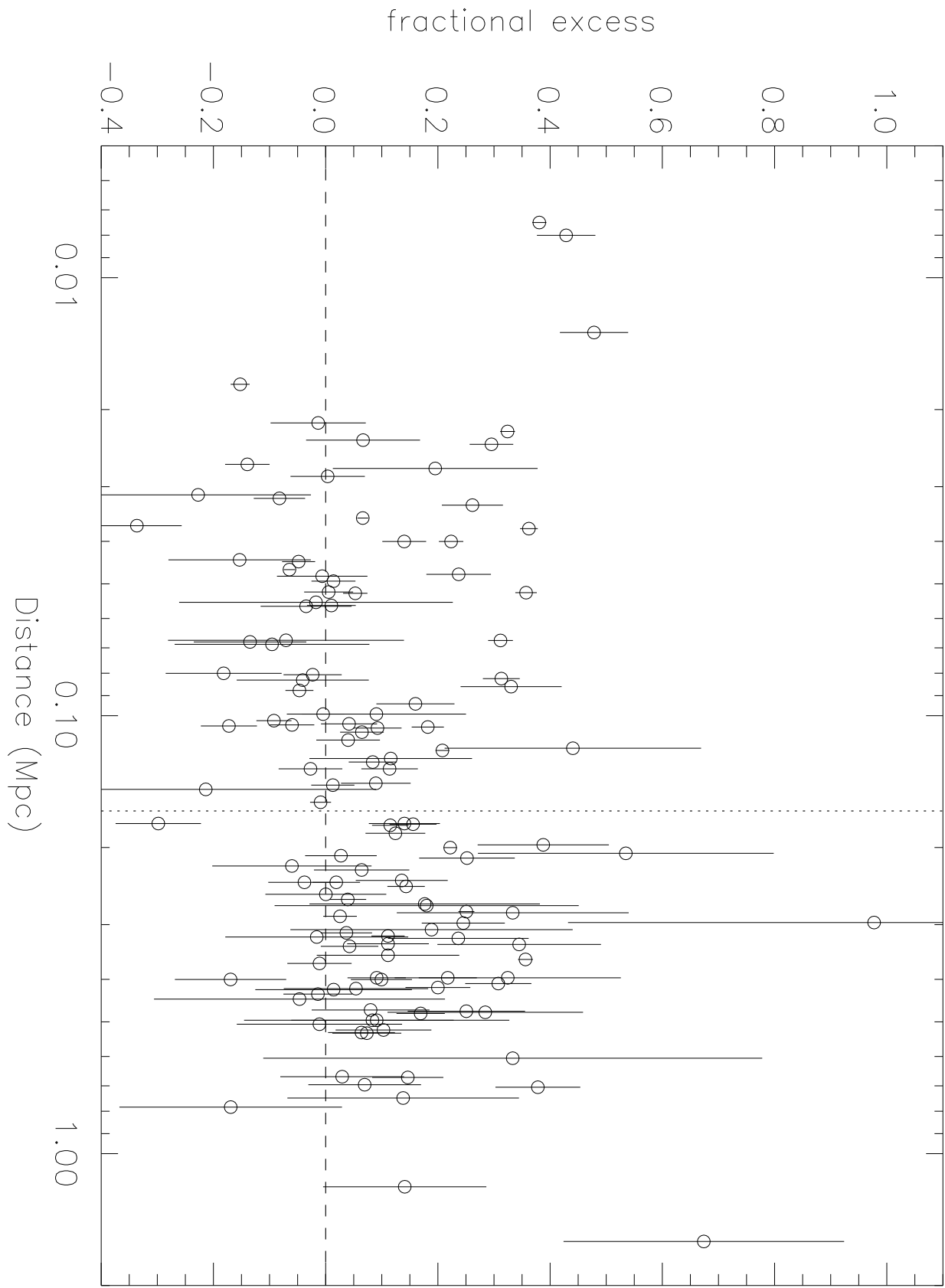
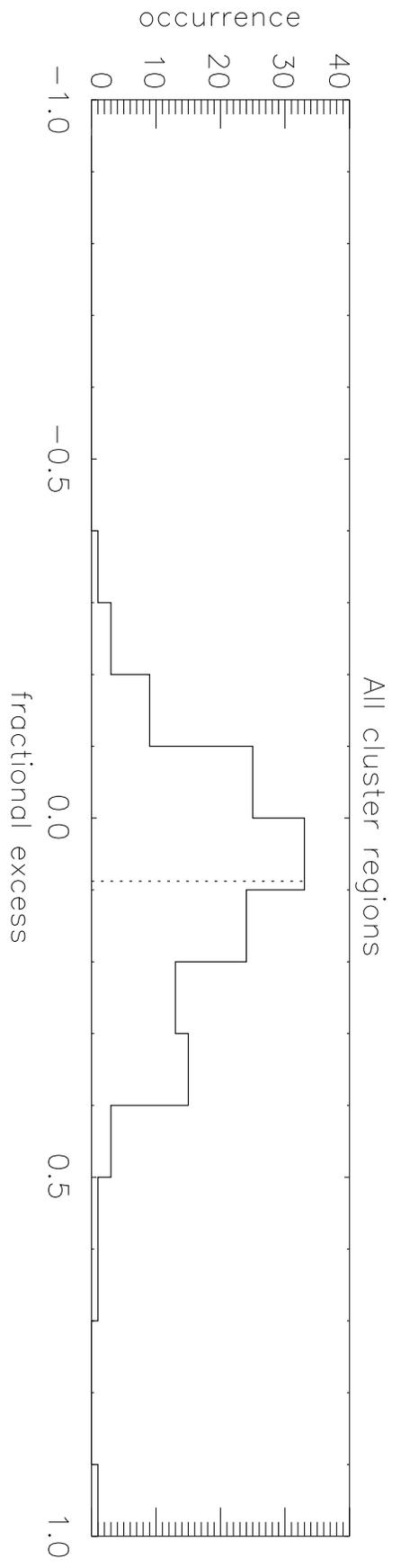
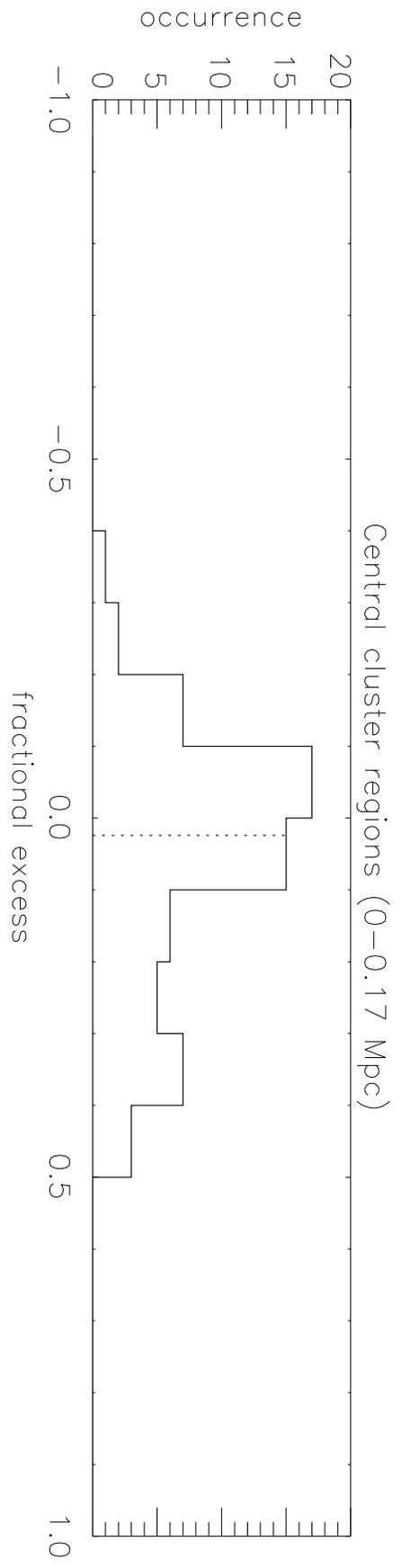
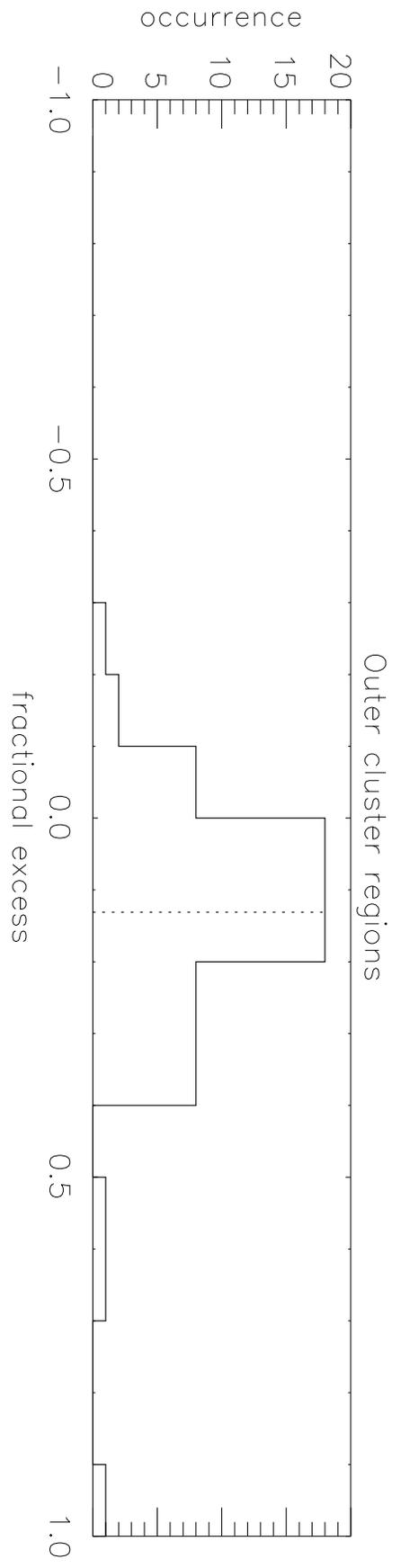
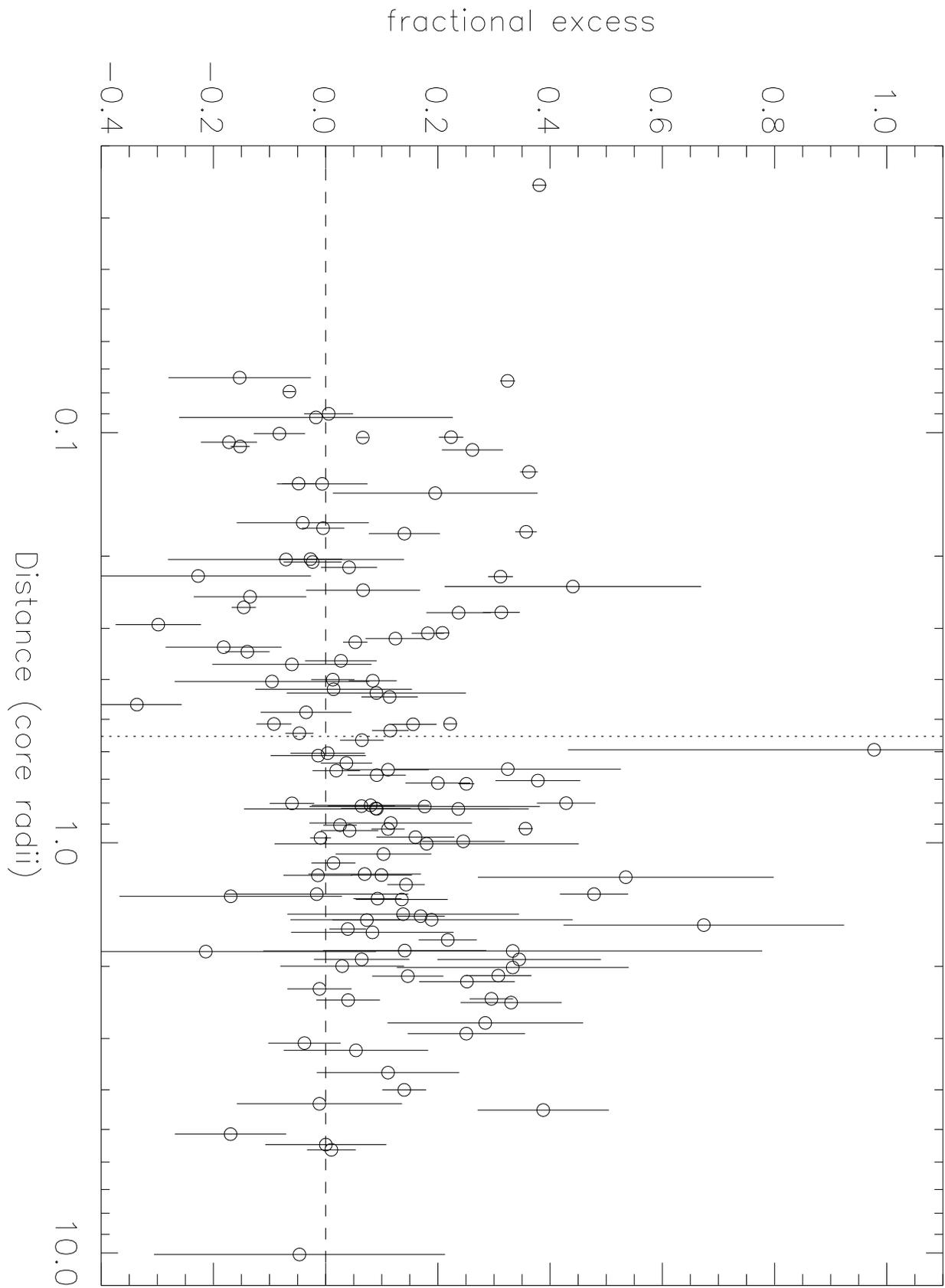
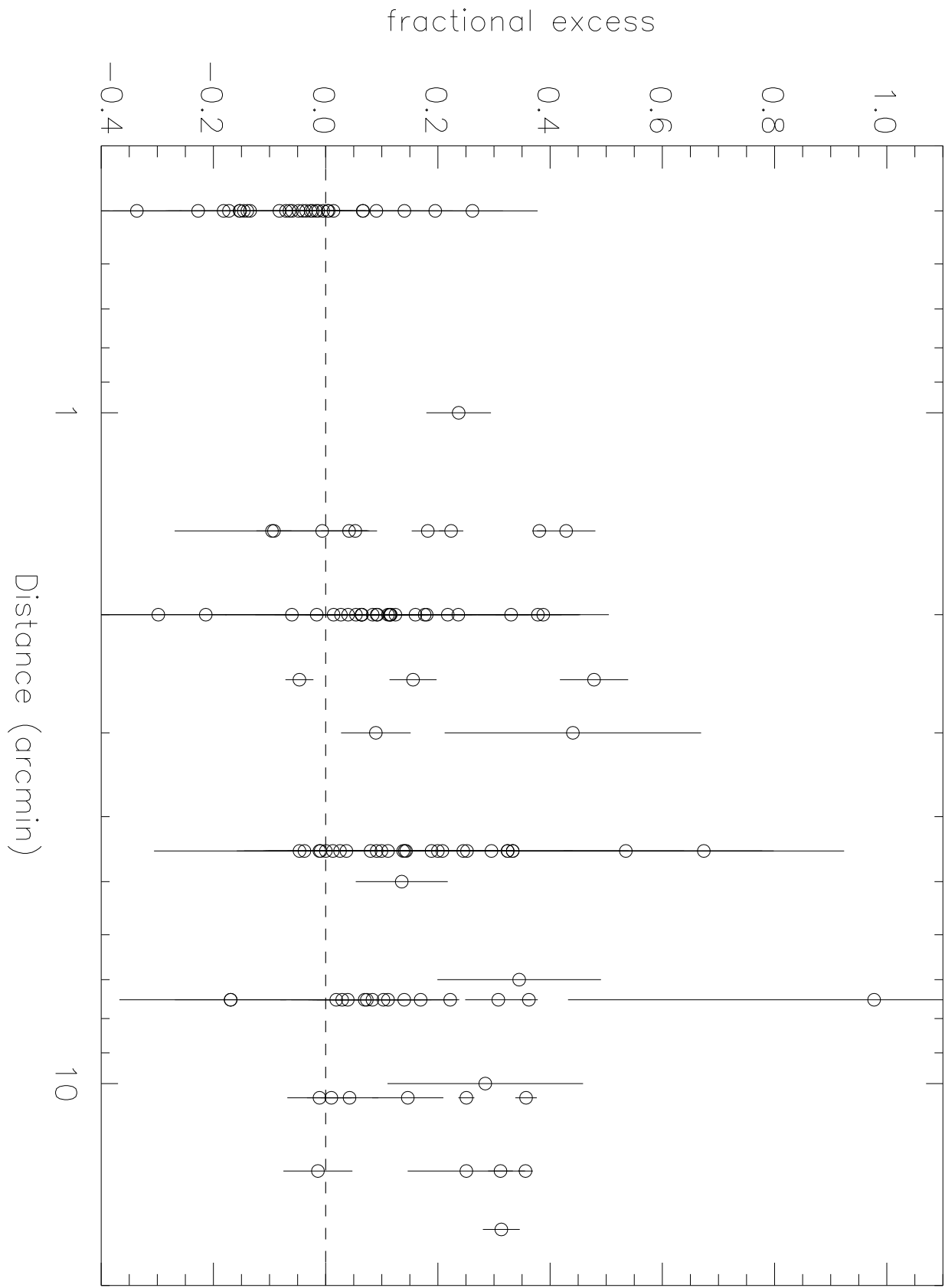
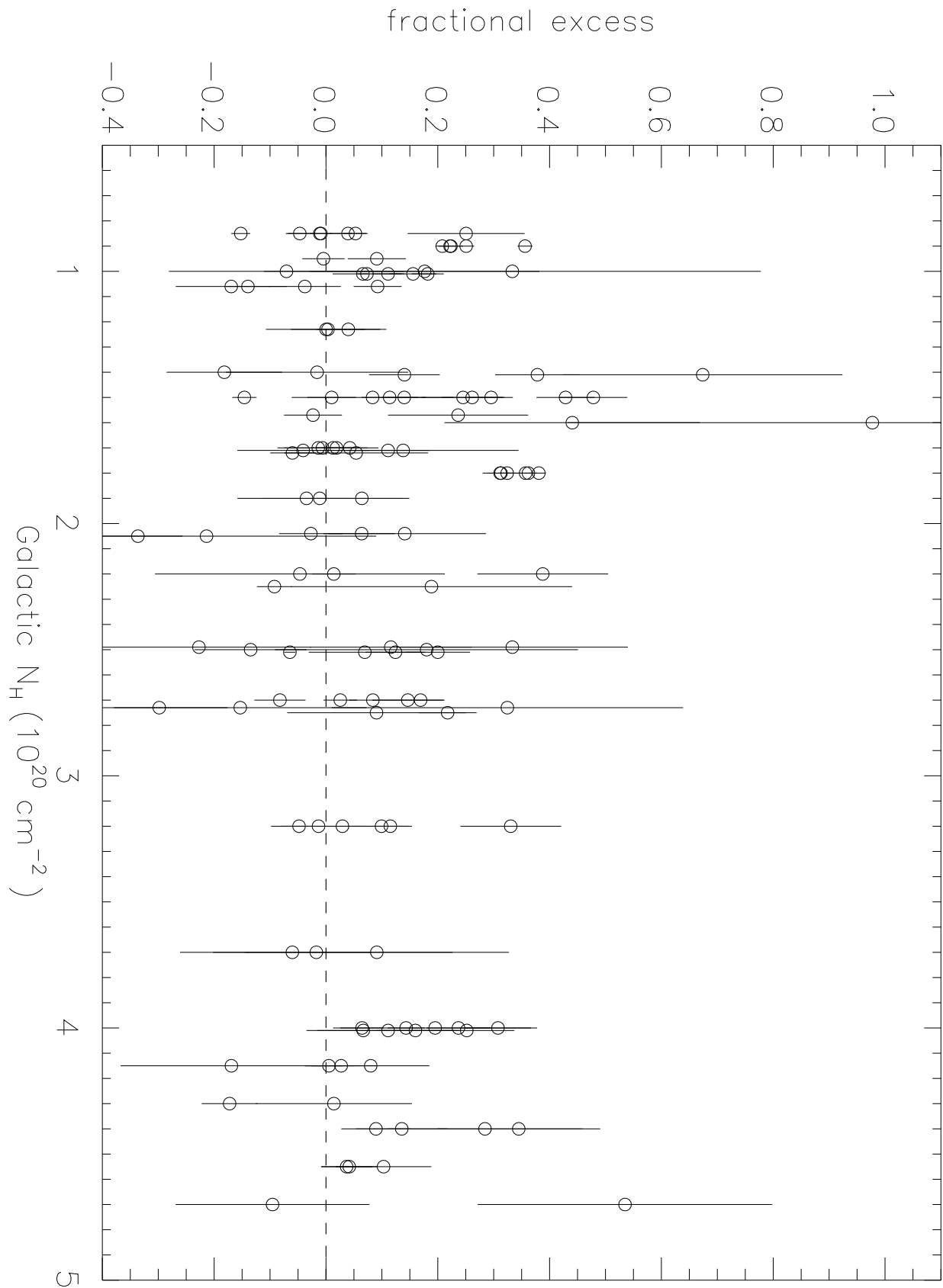
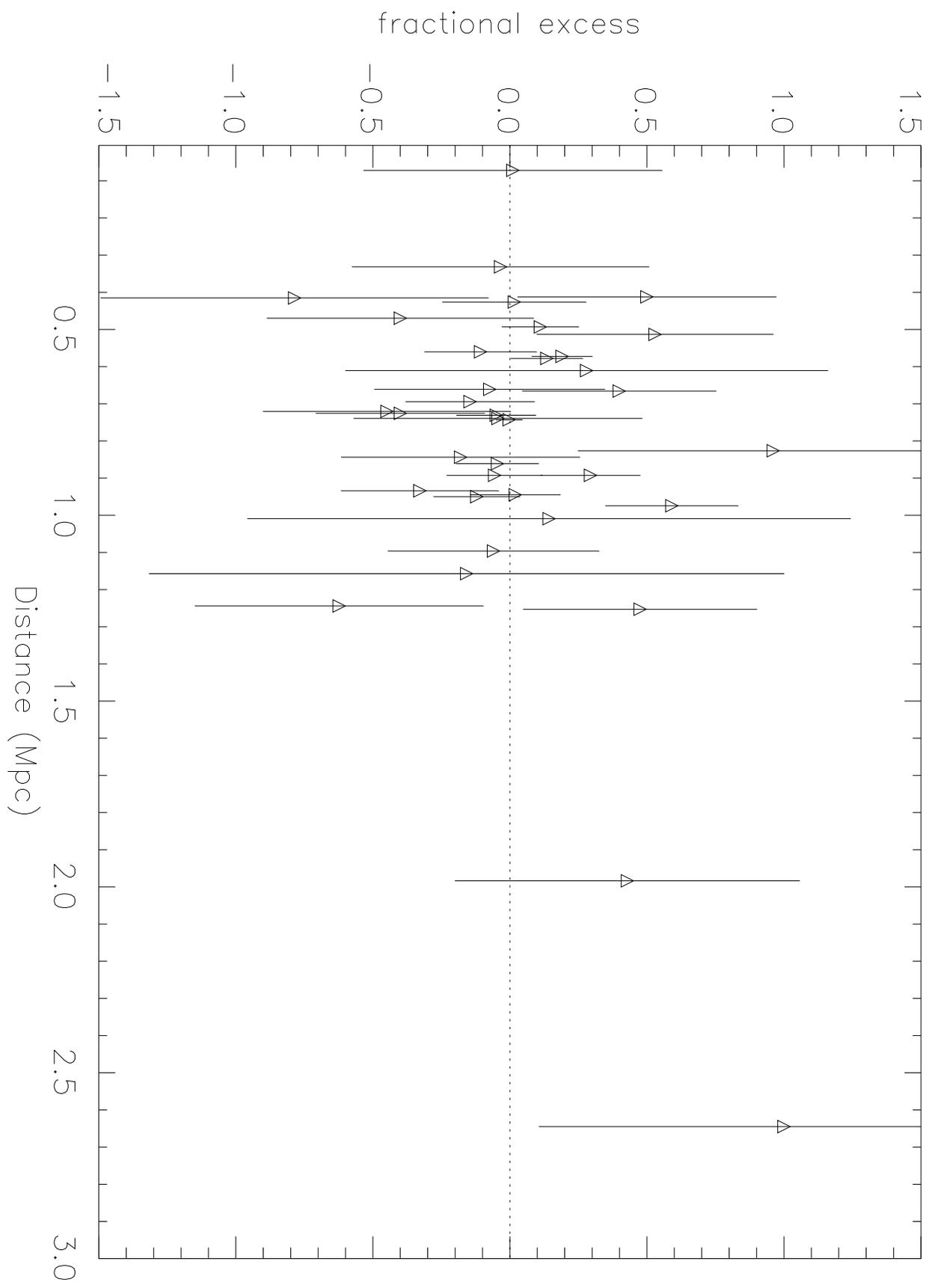
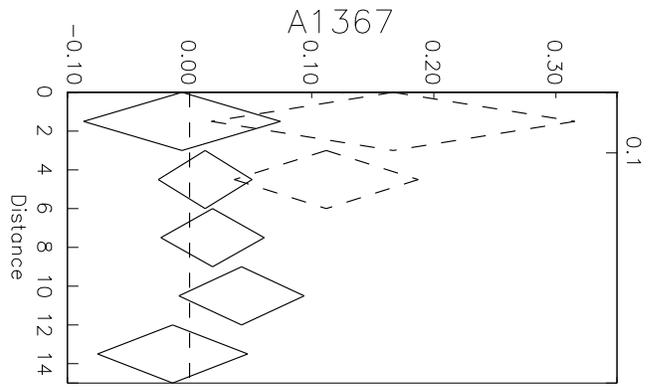
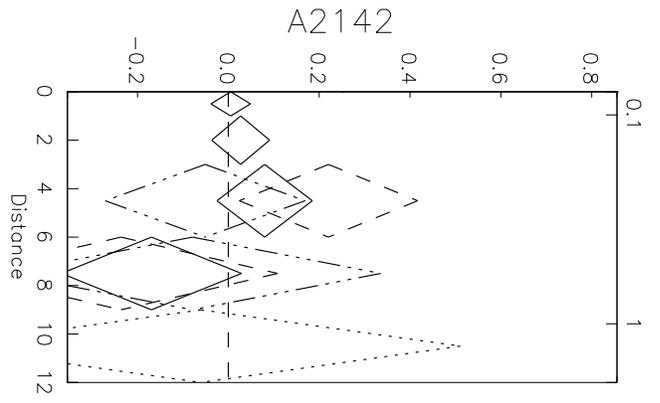
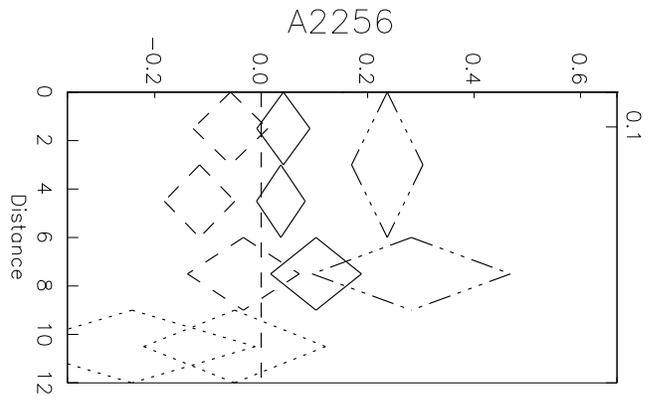
\end{document}